%% file: ntm.tex
\documentclass[twocolumn,aps]{revtex4}

\pdfoutput=1
\usepackage[T1]{fontenc}
\usepackage[utf8]{inputenc}
\usepackage[letterpaper]{geometry}
\usepackage{amsmath}
\usepackage{amssymb}
\usepackage{xfrac}
\usepackage{siunitx}
\usepackage{graphicx}
\usepackage{todonotes}
\usepackage{multirow}
\usepackage[caption=false]{subfig}

\makeatletter

\renewcommand{\arraystretch}{1.3} 

\usepackage[colorlinks=true,allcolors=blue]{hyperref}

\makeatother

\usepackage[american]{babel}

\begin{document}

\input{preamble}				
	\input{intro}				
	\input{methods}
	\input{overview}
	\input{analysis}
	\input{discussion}

	\input{acknowledge}
	\input{data}

	\bibliographystyle{apsrev4-1}
	\bibliography{ntm}

	\appendix
	\input{appendex}
\end{document}

%% file: preamble.tex
\title{Growing Neoclassical Tearing Modes Seeded via Transient-Induced-Multimode Interactions}

\author{E.C. Howell}
\email{ehowell@txcorp.com}
\affiliation{Tech-X Corporation, 5621 Arapahoe Avenue Suite A, Boulder, CO, 80303 USA}
\author{J.R. King}
\affiliation{Tech-X Corporation, 5621 Arapahoe Avenue Suite A, Boulder, CO, 80303 USA}
\author{J.D. Callen}
\affiliation{Department of Engineering Physics, University of Wisconsin - Madison, Madison, Wisconsin 53706 USA}
\author{R.J. La Haye}
\affiliation{General Atomics, P.O. 85608, San Diego, CA, 92186-5608, USA}
\author{R.S. Wilcox}
\affiliation{Oak Ridge National Laboratory, P.O. Box 2008, Oak Ridge, TN 37831, USA}
\author{S.E. Kruger}
\affiliation{Tech-X Corporation, 5621 Arapahoe Avenue Suite A, Boulder, CO, 80303 USA}

\begin{abstract}
\noindent Nonlinear extended MHD simulations demonstrating seeding of neoclassical tearing modes (NTMs) via MHD-transient-induced multimode interactions are presented. Simulations of NTMs are enabled by two recent NIMROD code developments: the implementation of heuristic neoclassical stresses and the application of transient magnetic perturbations (MPs) at the boundary. NTMs are driven unstable by the inherently pressure driven kinetic bootstrap current, which arises due to collisional viscosity between passing and trapped electrons. These simulations use heuristic closures that model the neoclassical electron and ion stresses. NTM growth requires a seed island, which is generated by a transiently applied MP in simulations. The capability is demonstrated using kinetic-based reconstructions with flow of a DIII-D ITER Baseline Scenario discharge [R.J. La Haye, et al., Proceedings IAEA FEC 2020]. The applied MP seeds a 2/1 NTM that grows in two phases: a slow growth phase followed by a faster robust growth phase like that observed experimentally. Additionally, an evolving sequence of higher order core modes are excited at first. Power transfer analysis shows that nonlinear interactions between the core modes and the 2/1 helps drive the initial slow growth. Once the induced 2/1 magnetic island reaches a critical width, the NTM transitions to faster robust growth which is well described by the nonlinear modified Rutherford equation. This work highlights the role of nonlinear mode coupling in seeding NTMs.
\end{abstract}




\date{\today}

\maketitle

%% file: intro.tex
\section{Introduction}

Neoclassical Tearing Modes (NTMs) are the leading root physics mechanism that causes disruptions in tokamaks \cite{deVries2011,Sweeney2016}. NTMs are macroscopic instabilities that arise due to a viscosity between trapped and passing particles. This viscous force gives rise to a pressure-gradient-driven bootstrap current; ultimately the helical perturbation of this current drives NTM growth \cite{QuCallen1985,Carrera1986,CallenIAEA86}. The existence of NTMs was first experimentally verified on TFTR \cite{ChangCallenPRL95}.

The standard theoretical tools for modeling NTM dynamics build upon the modified Rutherford equation (MRE) \cite{Rutherford1973,Sauter97}. The MRE analysis treats the evolution of thin isolated single-helicity islands. However, many open issues relating to the NTM evolution involve the interaction between multiple interacting modes or large islands with multiple helicities. Growing NTMs are often seeded by transient events such as edge localized modes (ELMs) or sawteeth \cite{ChangCallenPRL95,Sauter97}; the details of how these transients seed a NTM involves nonlinear interaction between multiple modes. For example, three-wave coupling between $m/n=1/1$ sawteeth with existing $4/3$ and $3/2$ islands has been identified as one mechanism for seeding $2/1$ NTMs in DIII-D ITER Baseline Scenario discharges (IBS) \cite{Bardoczi21} ($m$ and $n$ are the dominant poloidal and toroidal mode numbers). Disruptions also occur late in the evolution of a NTM, and the standard paradigm for island-induced disruptions involves multiple large overlapping islands \cite{Waddell79,Carreras80,Hu19}. Nonlinear simulations are well suited for handling these issues, bridging the gap between analytic theory and experiment.

However, nonlinear simulations of NTMs in realistic H-mode tokamaks are computationally challenging, and special considerations are required to make them tractable. NTMs evolve on slow transport times scales (typically on the order of \SI{100}{ms}). The time scale is long compared to the typical periods between ELMs and sawtooth crashes, and multiple ELM crashes and sawteeth are often observed during the evolution of a NTM. Nonlinear simulations of these transients are challenging and computationally expensive in their own right. For computational practicality, it is desirable to tailor realistic equilibria to avoid these transients. Simulations need to use realistic pedestal flow profiles in the edge to suppress ELMs, and the profile of the safety factor ($q$) needs to be tailored to keep the central $q_0 \gtrsim 1$ to avoid sawteeth. A self-consistent treatment of the neoclassical drives requires a five-dimensional kinetic treatment. While efforts are underway to couple fluid models with continuum kinetic closures \cite{Held15}, nonlinear simulations of the full NTM evolution is not yet tractable. For this reason heuristic closures are used to model the dominant neoclassical drives \cite{Gianakon2002}.

Multiple MHD codes have used heuristic models for neoclassical stresses. Gianakon first introduced the heuristic electron and ion stresses in  \textsc{nimrod} \cite{Gianakon2002} which were then used in simulations of the seeding of a $3/2$ NTM following a sawtooth crash in DIII-D \cite{Brennan2005}. Heuristic closures have also been implemented in the \textsc{XTOR} \cite{Maget2007} and \textsc{XTOR-2F} \cite{Mellet2013} MHD codes, and have been used to study $2/1$ NTMs in Tore Supra \cite{Maget2007} and JET \cite{Maget2010,Maget2011}. Recently, \textsc{JOREK} has used a heuristic ion stress, based off of Gianakon's closure, to model the neoclassical poloidal flow damping in RMP studies on ASDEX \cite{Meshcheriakov2019} and KSTAR \cite{SKKim2020}. The circular cross-section  reduced MHD code \textsc{TM1} has also implemented a heuristic bootstrap current, and has applied it to the problem such as radio-frequency (RF) stabilization of NTMs \cite{Yu04}.

This paper studies the seeding of a NTM via an applied transient external magnetic perturbation (MP) in nonlinear simulations. An $n=1$ perturbation containing a broad multiple-$m$ poloidal spectrum is applied at the boundary. The perturbation excites both a slowly growing $m/n = 2/1$ island and a sequence of higher order $m/n$ core modes. The focus here is on how nonlinear interactions between core modes helps sustain the $2/1$ growth until it reaches a critical amplitude for robust MRE growth. This applied MP can be viewed as a surrogate model for ELM seeding of a NTM. ELMs result from the nonlinear evolution of multiple moderate $n$ linearly unstable peeling-ballooning modes. The $n$-th and $n$-th+1 ($\sim 6-8$) modes nonlinearly interact to produce an $n=1$ perturbation \cite{Pankin2007}.

%
The outline of the paper is as follows. Section \ref{sec:methods} discusses NIMROD model developments \cite{NIMRODJCP} that are needed for this study.
Section \ref{sec:overview} presents an overview of a simulation of a growing $2/1$ NTM in a classically stable ITER-relevant DIII-D equilibrium. Section \ref{sec:analysis} investigates the nonlinear multi-mode interactions that drive the $2/1$ growth using a Fourier mode energy transfer analysis which is derrived in Appendix \ref{apx:PowerTransfer}. Finally, section \ref{sec:discussion} concludes with a discussion of these results.

%% file: methods.tex
\section{Model Development}\label{sec:methods}

This section introduces two developments that enable NTM modeling in \textsc{nimrod} \cite{NIMRODJCP}. The first development is the use of heuristic closures that model the neoclassical stresses. The second development is the application of an external transient magnetic perturbation which generates the seed island.

\subsection{Heuristic Neoclassical Stress Closures}
\textsc{nimrod} simulates the nonlinear evolution of the primitive field variables using extended MHD models. This work incorporates neoclassical effects by including a heuristic neoclassical ion stress in the plasma momentum equation and a neoclassical electron stress in Ohm's law. Here the stresses are incorporated into the resistive MHD model:

\begin{align}
&\frac{D n}{D t} = -n \nabla \cdot \vec v + D_n\nabla^2 n, \label{eqn:cont}\\
&\rho \frac{D \vec v}{D t} = \vec J \times \vec B - \nabla p  - \nabla \cdot \vec{\vec{\pi}}_{i}, \label{eqn:mom}\\
&\frac{3}{2}n\frac{D T_\alpha}{D t}  = -p_\alpha \nabla \cdot v - \nabla \cdot \vec q_\alpha, \label{eqn:temp}\\
&\frac{\partial \vec B}{\partial t} = - \nabla \times \vec E +\kappa_{B}\nabla \nabla \cdot \vec B,\label{eqn:induction}\\
&E = -\vec v \times \vec B + \eta \vec J -\frac{1}{ne}\nabla\cdot \vec{\vec{\pi}}_{e}, \label{eqn:ohms}\\
&\nabla \times \vec B = \mu_0 \vec J. \label{eqn:Amp}
\end{align}

\noindent The ion stress tensor, $\vec{\vec{\pi}}_{i}$, includes both the classical stress and the neoclassical stress. The electron stress tensor, $\vec{\vec{\pi}}_{e}$, only includes the neoclassical electron stress. The fluid variables $n\left(\rho\right)$, $\vec v$, and $p$ are the fluid number (mass) density, center of mass flow velocity, and the pressure. The material derivative
\begin{equation}\label{eqn:material_d}
\frac{D }{D t}= \frac{\partial}{\partial t} + \vec v \cdot \nabla
\end{equation}
includes both the Eulerian time derivative and advection due to the center of mass flow. The thermodynamic variables $T$ and $\vec q$ are the temperature and heat flux. The subscript $\alpha$ denotes the species. The electromagnetic variables $\vec E$, $\vec B$ and $\vec J$ are the electric field, magnetic field, and current density. The permeability of free space is $\mu_0$, and the electrical resistivity is $\eta$. Simulations include an artificial particle diffusivity, $D_n$, in Eq. (\ref{eqn:cont}) to damp small-scale density fluctuations, and a magnetic-divergence diffusivity $\kappa_{B}$ in Eq. (\ref{eqn:induction}) to control numerical magnetic field divergence errors.

The heuristic force resulting from the neoclassical ion stress is modeled as
\begin{equation}\label{eqn:neoi}
\nabla\cdot \vec{\vec{\pi}}_{i} = \mu_i n m_i \left<B_{eq}^2\right> \frac{\left(\vec v-\vec v_{eq}\right)\cdot \vec e_\Theta}{\left(\vec B_{eq}\cdot \vec e_\Theta\right)^2}\vec e_\Theta,
\end{equation}
and the heuristic force resulting from the neoclassical electron stress is \cite{Gianakon2002}
\begin{equation}\label{eqn:neoe}
\nabla\cdot \vec{\vec{\pi}}_{e} = -\mu_e \frac{m_e}{e} \left<B_{eq}^2\right> \frac{\left(\vec J-\vec J_{eq}\right)\cdot \vec e_\Theta}{\left(\vec B_{eq}\cdot \vec e_\Theta\right)^2}\vec e_\Theta.
\end{equation}

\noindent Here $\mu_\alpha$ is the $\alpha$ species neoclassical poloidal flow damping rate, $e$ is the electron charge, brackets $\left< \dots \right>$ indicate a flux surface average (FSA), equilibrium quantities are indicated using subscript ``eq'', and $\vec e_\Theta$ is the flux-aligned poloidal basis vector.

Equations (\ref{eqn:neoi}) and (\ref{eqn:neoe}) are simplifications of the general species viscous force denisty
\begin{equation}\label{eqn:neoa}
\nabla\cdot \vec{\vec{\pi}}_{\alpha} = \mu_\alpha n_\alpha m_\alpha \left<B_{eq}^2\right> \frac{\vec v_\alpha \cdot \vec e_\Theta}{\left(\vec B_{eq}\cdot \vec e_\Theta\right)^2}\vec e_\Theta
\end{equation}
introduced in reference \cite{Gianakon2002}. The ion force assumes that the ion flow velocity is the fluid flow velocity, $\vec v = \vec v_i$. This assumption is valid to order $O\left(m_e/m_i\right)$. The electron stress assumes that the poloidal electron flow velocity is $\vec v_e \cdot \vec e_\Theta = -\frac{1}{ne}\vec J \cdot \vec e_\Theta$, and is valid in the limit of strong ion poloidal flow damping: $\mu_i \gg 1$. As written here, the expressions assume an effective ion charge $Z_{eff} = 1$, and thus $n=n_i=n_e$.

The closures model the neoclassical bootstrap current, neoclassical ion poloidal flow damping, and the neoclassical ion polarization current enhancement in a way that makes simulations tractable. The closures only depend on variables that are readily available in the fluid simulation. The factor $\left<B_{eq}^2\right>$ is calculated when the initial equilibrium is generated. The poloidal basis vector $\vec e_\Theta$ is proportional to the equilibrium poloidal magnetic field, and the constant of proportionality cancels in Eqs. (\ref{eqn:neoi}) and (\ref{eqn:neoe}). The subtraction of equilibrium flow in Eq. (\ref{eqn:neoi}) and the equilibrium current density in Eq. (\ref{eqn:neoe})  maintains these profiles against decay.



\subsection{Seed Island Generation via Transient Magnetic Perturbations}
Our interest is in studying NTMs in classically stable conditions. These NTMs are linearly stable, and require a seed island for growth. This complicates simulations in realistic shaped equilibria, where calculating a sufficiently accurate seed perturbation is nontrivial. To address this, a technique to generate the seed using transient forced magnetic reconnection \cite{Hahm1985,Beidler2018} is developed. This method is inspired by the fact that MHD transients frequently seed NTMs in experiments \cite{LaHaye2006}. Resonant components of the perturbation tear the magnetic field at the resonant rational surface, and a sufficiently large perturbation will generate a growing seed island.

Simulations apply a transient magnetic perturbation at the computational boundary of the form
\begin{equation}\label{eqn:MpPulse}
\hat n \cdot \vec B\left(\vec x,t\right) = \hat n \cdot \vec B_{mp}\left(\vec x \right) \times \text{Env}\left(t\right)\times \text{exp}\left(i\Omega t\right).
\end{equation}
Here $\hat n \cdot \vec B\left(\vec x,t\right)$ is the normal component of the time varying magnetic perturbation, $\vec B_{mp}\left(\vec x \right)$ characterizes the spatial variations of the magnetic perturbation, $\text{Env}\left(t\right)$ is the time envelope of the perturbation, and the exponential factor modulates the perturbation with angular frequency $\Omega$.

A bump function envelope is used:

\begin{equation}\label{eqn:pulse}
\text{Env}=\begin{cases}
E_0 \times \text{exp}\left(-\frac{1}{1-\tau^2} \right) & \text{for } \tau \in \left[-1,1\right]\\
0 &\text{otherwise},
\end{cases}
\end{equation}
\noindent where $\tau= \left(t- t_o - t_w\right)/t_w $. The parameter $E_0$ is a scale factor that varies the amplitude of the applied pulse, $t_w$ is the pulse half-width, and $t_o$ is a time offset. The bump function pulse is numerically advantageous because it has compact support (it is only nonzero in a finite domain) and it is smooth (derivatives are continuous to all orders). The applied pulse has a peak amplitude of $E_0/e^1$ at time $t=t_w+t_o$.

The factor $\text{exp}\left(i\Omega t\right)$ is used to change the frequency modulation of the applied MP to maximize the plasma response. Linear theory predicts that flow screening is minimized and the largest plasma response occurs when the applied RMP is almost in phase with the rotation of the resonant mode \cite{Fitzpatrick91}.


\begin{figure}[h]
\centering
\includegraphics[width=0.5\textwidth]{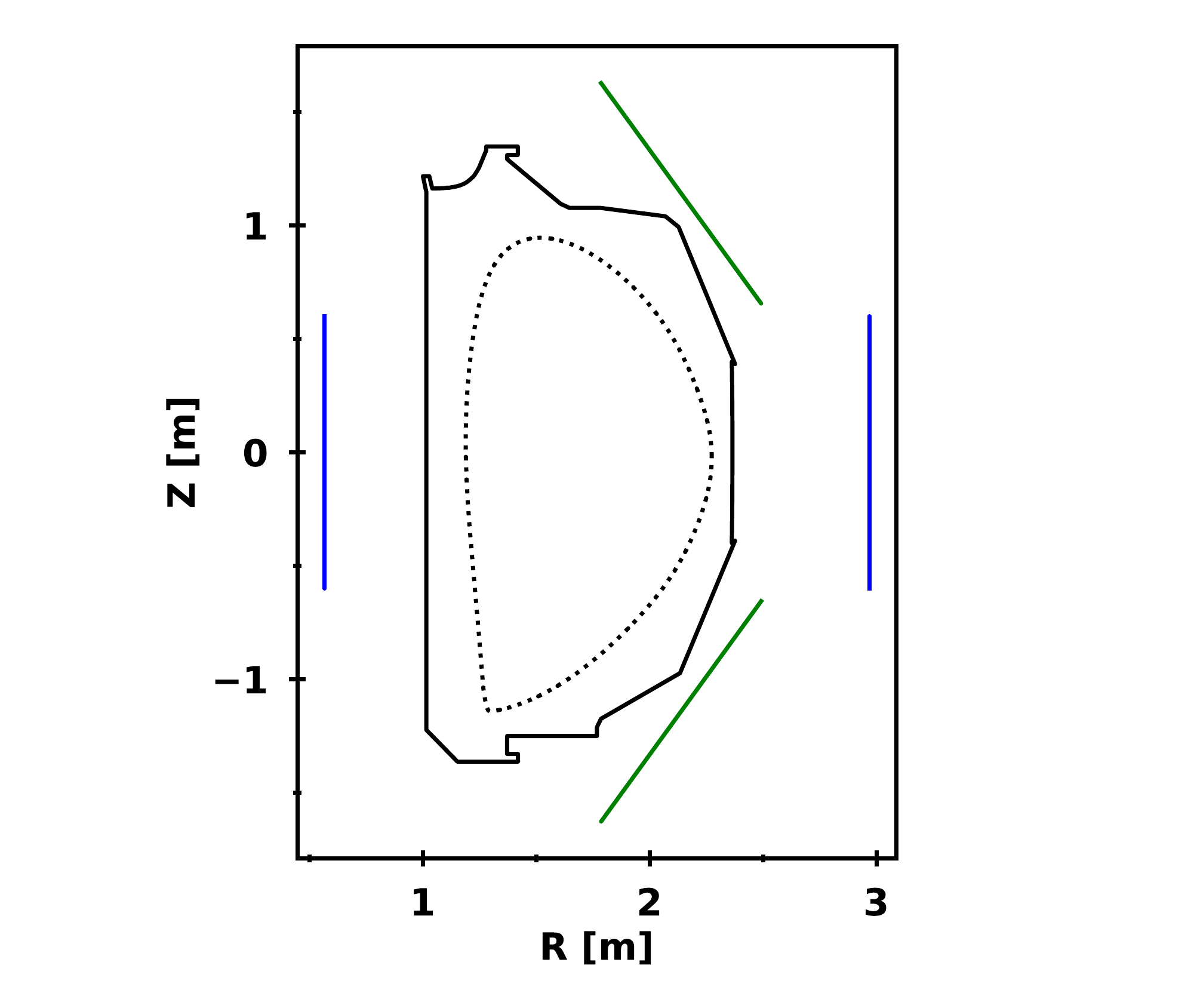}
\caption{
The poloidal location of the circular coils used to generate the applied magnetic perturbation are indicated by the blue and green lines. The alternating blue-green lines emphasize the alternating direction of the current in the coils. There are six arrays of coils spaced evenly in the toroidal direction. The DIII-D limiter location is indicated by the solid line, and the computational boundary, which corresponds to the reconstructed equilibrium separatrix, is indicated by the dotted line.
}
\label{fig:coils}
\end{figure}

The poloidal and toroidal spectrum of the applied MP is determined by the factor $\vec B_{mp}\left(\vec x \right)$. Here $\vec B_{mp}$ is calculated using a planar coil array consisting of six toroidal sets of circular coils. Each set consists of four coils arranged poloidally around the computational domain as shown in Fig. \ref{fig:coils}. The poloidal orientation of the coil is motivated by considering cylindrical geometry, where poloidal coils arranged in a cross with the current flowing in alternating directions in adjacent coils will produce a geometric $m=2$ perturbation. However, in toroidal geometry with plasma shaping this configuration produces a response with a broad poloidal spectrum. Vacuum response calculations are used to further refine the coil locations and coil pitches to increase the resonant 2/1 helical flux (defined in SEC. \ref{sec:overview}) relative to the resonant 3/1 and 4/1 helical fluxes. To further tailor the applied MP, $\vec B_{mp}\left(\vec x \right)$ is Fourier transformed in the toroidal direction, and only the $n=1$ Fourier mode is applied in the simulations presented here.


%% file: overview.tex
\section{Simulation Overview}\label{sec:overview}

This section presents a high level overview of the simulation results. First, the details of the equilibrium reconstruction are discussed. Then the qualitative dynamics of the simulation are presented.

\subsection{Experimental Equilibrium}

Experimental equilibrium reconstructions are used to model NTM dynamics in realistic conditions relevant to DIII-D, ITER, and future burning plasma tokamaks. Simulations use a kinetic reconstruction of an experimental DIII-D ITER Baseline Scenario (IBS) discharge \cite{LaHaye21, Callen21}. The reconstruction is of an ELMing H-mode discharge with ITER similar shape, $\beta_n \sim 1.8$, and $q_{95} \sim 3.4$.

\begin{figure}[h]
\centering
\includegraphics[width=0.5\textwidth]{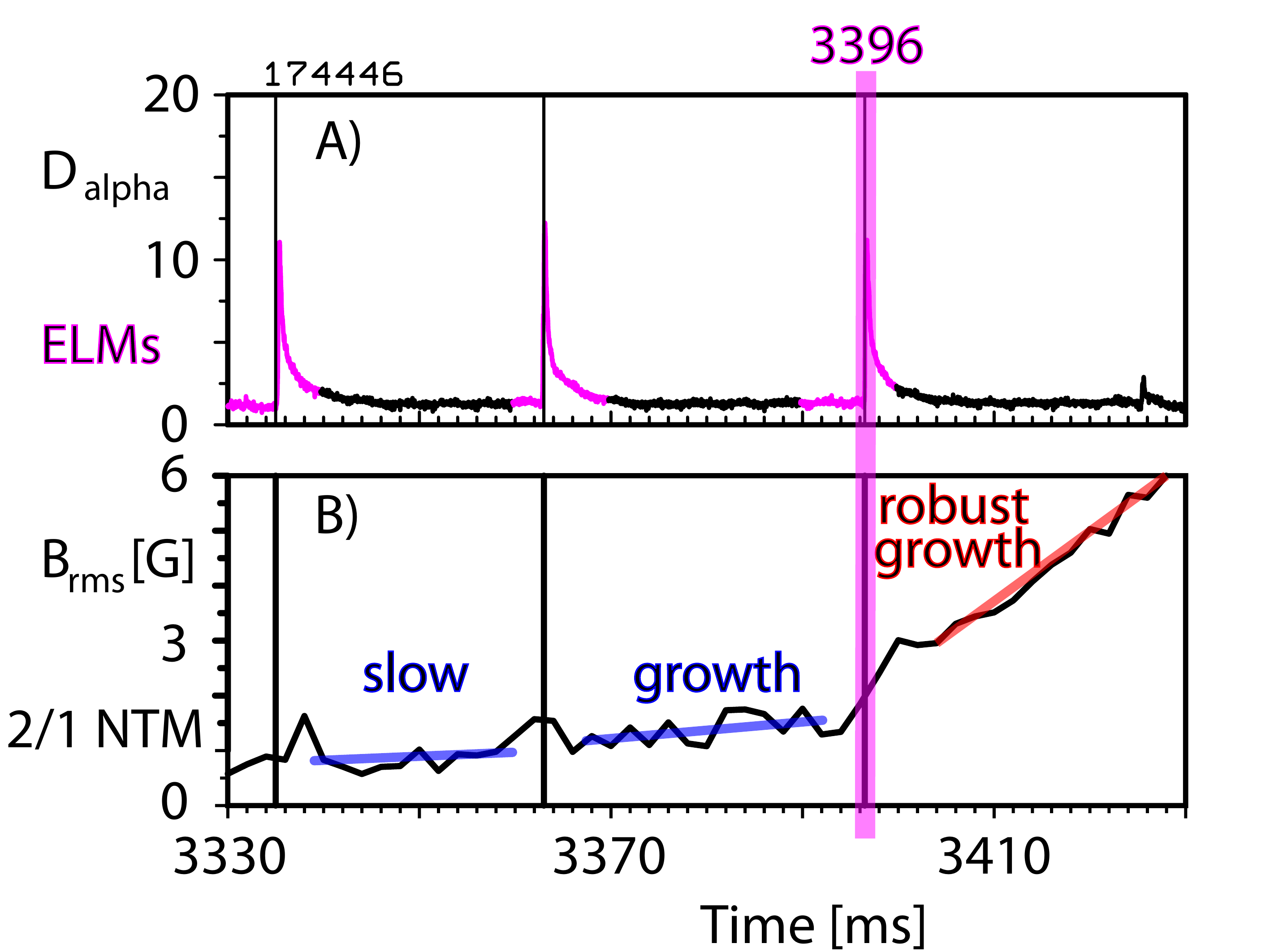}
\caption{Experimental time traces of the $\text{D}_\text{alpha}$ signal, and $n=1$ $B_\text{rms}$ for the DIII-D reference discharge used in this study. An ELM at \SI{3396}{ms} excites a robustly growing $2/1$ NTM. Simulations use kinetic reconstruction of conditions at \SI{3390}{ms}.}
\label{fig:exp_time}
\end{figure}

The kinetic reconstruction of the equilibrium is for the DIII-D IBS discharge 174446 at \SI{3390}{ms}, just before an ELM seeds a robustly growing NTM. Figure \ref{fig:exp_time} show the experimental time traces of the reference discharge in a \SI{100}{ms} window around the time of interest. The $\text{D}_\text{alpha}$ signal shows a series of ELMs around \SI{3335}{ms}, \SI{3363}{ms}, and \SI{3396}{ms}. A small slowly growing $n=1$ signal is observed following the first two ELMs. Then the ELM at \SI{3396}{ms} triggers a larger robustly growing $2/1$ NTM.

This discharge is a part of an experimental campaign to study seeding of $2/1$ NTMs. The diagnostics in this discharge were aligned to provide high resolution measurements in the vicinity of the $q=2$ surface. In addition, the discharges use fast (\SI{1}{ms}) CER measurements \cite{Chrystal16} to provide time resolved measurements of the toroidal and poloidal flows. These high resolution measurement enable high fidelity kinetic reconstructions which are needed for simulation.

\begin{figure}[h]
\centering
\includegraphics[width=0.5\textwidth]{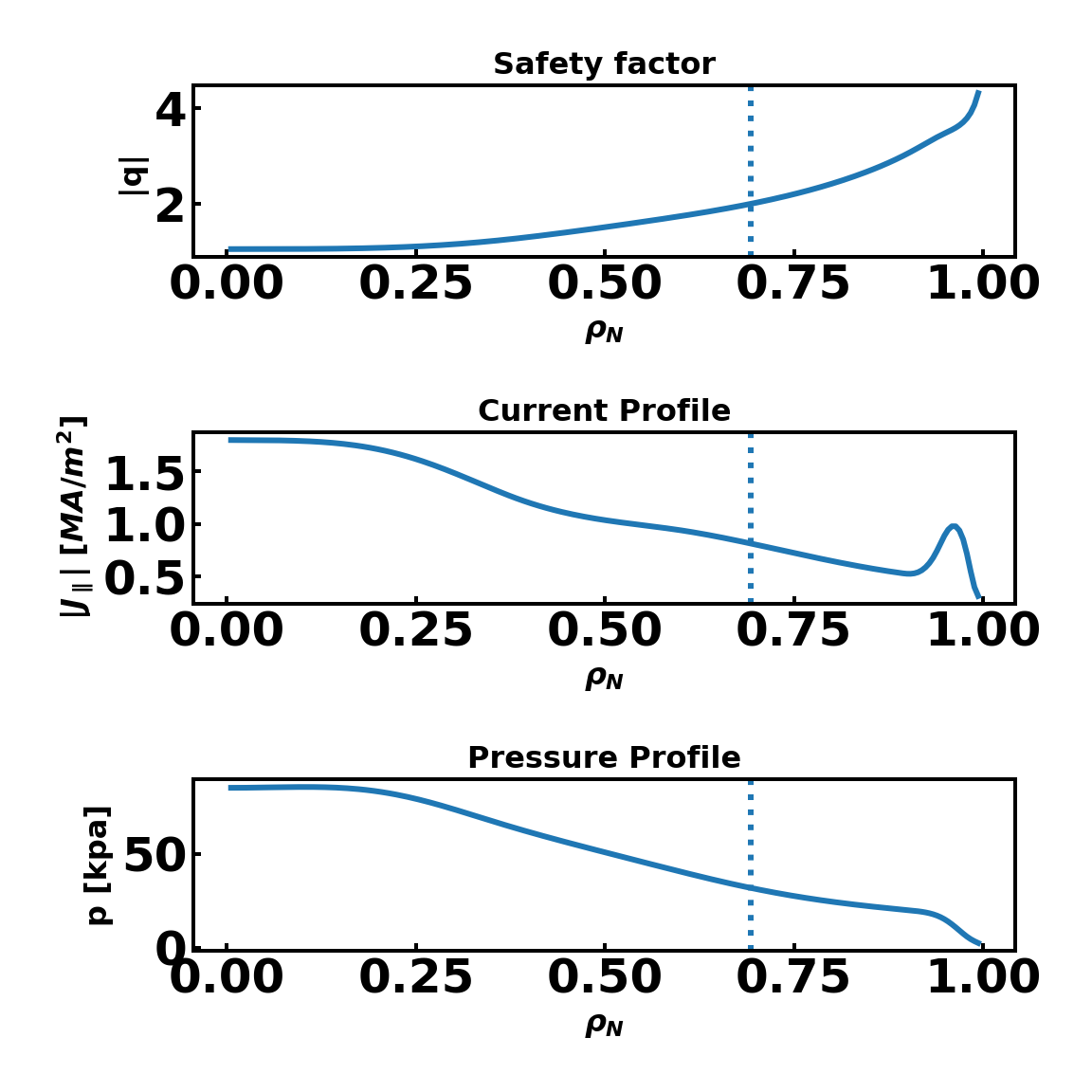}
\caption{The equilibrium safety factor ($q$), flux-surfaced-averaged parallel
current density $\langle J_\parallel \rangle$, and pressure ($p$) profiles used in the simulation. The
vertical dotted line indicates the location of the $q=2$ surface.}
\label{fig:eqprof}
\end{figure}

Kinetic EFIT \cite{EFIT} reconstructions are generated using the OMFIT Framework \cite{OMFIT,logan18}. In addition to the standard diagnostics used in reconstructions, kinetic reconstructions constrain the pressure profile using calculated fast ion pressures and the reconstructions constrain the current profile using calculated bootstrap current. Enabled by the time resolved measurements, the reconstruction only use data in the window preceding the ELM crash at \SI{3396}{ms} to accurately represent conditions immediately prior to the NTM onset (the standard method averages over multiple ELM cycles).

The reconstructed safety factor, parallel current, and pressure profiles used in the simulation are shown in Fig. \ref{fig:eqprof}. The safety factor at the magnetic axis is artificially increased to be slightly greater than 1 ($q_0>1$), to stabilize the $1/1$ internal kink and avoid sawteeth. This creates a region of low magnetic shear near the axis, and impacts the stability of core modes in this region. Care is taken to preserve the safety factor profile in the vicinity of the $q=2$ surface.

\begin{figure}[h]
\centering
\includegraphics[width=0.5\textwidth]{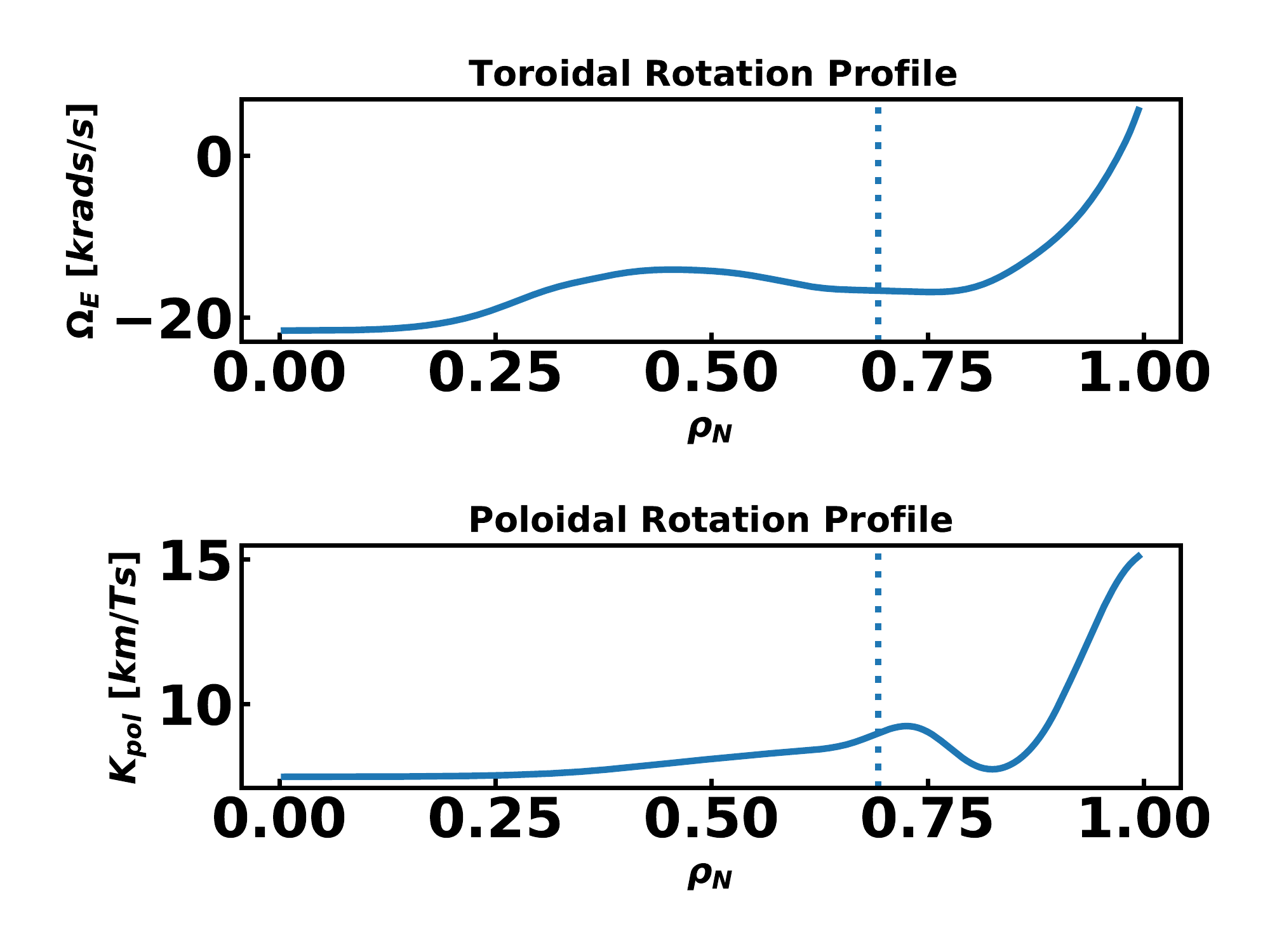}
\caption{The equilibrium toroidal ($\Omega_E$) and poloidal ($K_{pol}$) rotation
profiles used in the simulation. The gradients in the rotation profiles in the
pedestal ($\rho_N \gtrsim 0.9$) produce flow shear which stabilizes the peeling
ballooning modes. The vertical dotted line indicates the location of the $q=2$
surface. }
\label{fig:eqrot}
\end{figure}

Simulations also include the reconstructed toroidal and poloidal flow profiles to accurately model the interaction between different modes. The inclusion of the flow shear in the pedestal is also needed to stabilize peeling-ballooning modes. The equilibrium flow is given by the expression
\begin{equation}\label{eqn:veq}
v_{eq} = K\left(\rho_N\right) \vec B_{eq} + \Omega_E\left(\rho_N\right) R^2 \nabla \phi,
\end{equation}
where $K\left(\rho_N\right)$ and $\Omega_E\left(\rho_N\right)$, shown in Fig. \ref{fig:eqrot}, are the reconstructed poloidal and toroidal rotation profiles. The diamagnetic flow is neglected, consistent with the single fluid MHD model used in this study.

The simulations use the reconstructed electron density, electron temperature, and total pressure profiles. The equilibrium ion temperature is then calculated using the definition of the pressure $p=n(T_e + T_i)$. The experimental reconstructions include the fast particle pressure and impurities, which are not in the MHD model. As a result, there is not enough flexibility in the MHD model to match all of the reconstructed electron density, electron temperature, total pressure, and the ion temperature profiles. The simulated ion temperature differs from the reconstructed ion temperature. The ion temperature does not directly impact any of the transport coefficients in the model used; this difference will have a minimal impact on dynamics.

A Spitzer resistivity profile based on the equilibrium electron temperature is used $\eta = \eta_0 T_e^{-3/2}$, but the neoclassical electron and ion damping rates are uniform. The resistivity and damping coefficients are artificially enhanced for numerical convenience; however the ratio $\frac{\mu_e}{\nu_{ei}+\mu_e}=0.55$ at the $q=2$ surface is chosen to be close to the experimental value. This ratio characterizes the neoclassical drive relative to the classical resistive drive. The simulated collision frequency is calculated using $\nu_{ei} = \eta\frac{ne^2}{m_e}$.

\begin{table}[h]
  \begin{tabular}{|c|c|c|}
    \hline
     & Simulation & Experiment \\
    \hline
    Lundquist Number & $2.5 \times 10^6$ & $7.9 \times 10^6$\\
    \hline
    Prandtl Number & $23$ & $11$ \\
    \hline
    $\left(\chi_\parallel / \chi_\perp\right)^{1/4}$ & $100$ & $260$ \\
    \hline
    $\mu_e$ & $8 \times 10^5 [s^{-1}]$ & $1.3\times 10^5[s^{-1}]$\\
    \hline
    $\mu_i$ & $1 \times 10^3 [s^{-1}]$ & $1.4\times 10^3[s^{-1}]$\\
    \hline
    $\mu_e / \left(\nu_{ei} + \mu_e\right)$ & $0.55$ & $0.43$\\
    \hline
  \end{tabular}
\caption{Simulation parameters evaluated at the $2/1$ surface, and experimental parameters for comparison.}
\label{tab:simParam}
\end{table}

Table \ref{tab:simParam} shows some of the relevant simulation parameters calculated at the $2/1$ surface. The corresponding experimental parameters are also shown for comparison. In terms of the normalized parameters the simulations parameters are within a factor of 2 or 6 of the experiment parameters.

\subsection{Simulation Phenomenology}

\begin{figure}[!h]
\centering
\includegraphics[width=0.5\textwidth]{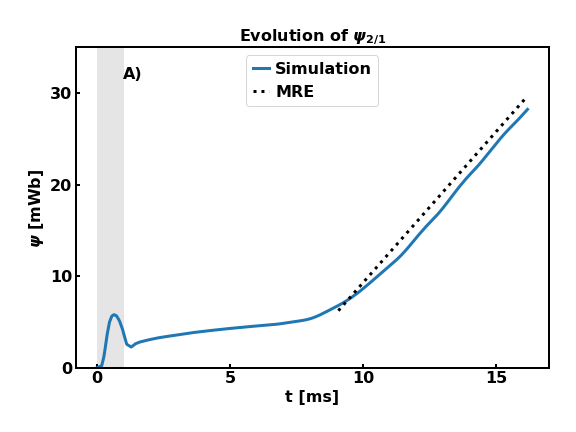}
\newline
\includegraphics[width=0.5\textwidth]{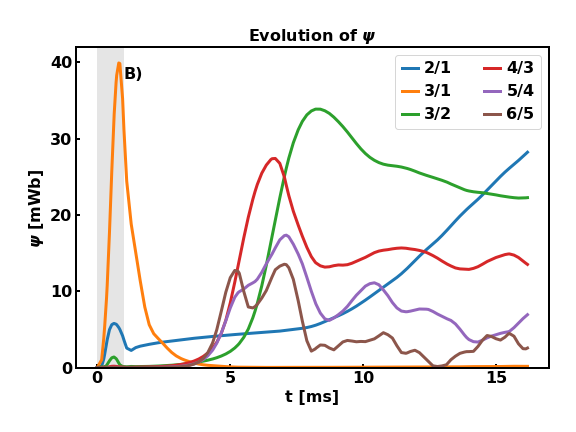}
\caption{Evolution of helical resonant flux calculated at the respective resonant surfaces following the application of a \SI{1}{ms} magnetic pulse. A) Shows the evolution of the $2/1$ resonant flux for clarity. B) Shows the evolution of multiple low order resonant helical fluxes.} \label{fig:psi_evo}
\end{figure}

This section provides an overview of the evolution of the NTM following
the application of a \SI{1}{ms} magnetic perturbation. The amplitude of the resonant helical magnetic flux $\psi_{m,n}$, is used throughout the discussion as the measure of the magnetic island width. The helical flux is defined as

\begin{equation}\label{eqn:helflux}
\psi_{m,n} = \oint \mathcal{J}\tilde B \cdot \nabla \rho \exp \left(in\phi -im\Theta \right)d\Theta d\phi,
\end{equation}

\noindent where $\left(\rho, \Theta, \phi\right)$ are toroidally axisymmetric flux-aligned coordinates and the Jacobian is

\begin{equation}
\mathcal{J}^{-1}=\nabla \rho \cdot \left(\nabla \Theta \times \nabla \Phi \right).
\end{equation}

The definition of the helical flux is inspired by the commonly used SURFMN calculation of $\widetilde b_r$ \cite{Schaffer2008}; however, a different normalization is used \cite{Park08}. Equation (\ref{eqn:helflux}) has the advantage of being independent of the choice of radial coordinate variable $\rho$.  In the simulation the two quantities are related by the expression $\widetilde b_r [G]=0.544\left|\psi_{2,1} \right|[mWb]$ for the $2/1$ island.

Assuming a thin single-helicity island, the island width in units of flux is
\begin{equation}
w = \frac{4}{2\pi}\sqrt{\frac{q}{q'}\frac{\left|\psi_{m,n} \right|}{m}},
\end{equation}
where $q'$ is the derivative of $q$ with respect to the equilibrium poloidal flux\cite{Schaffer2008}. The width of a $2/1$ island in can be calculated from the helical resonant flux using the expression $w[cm] = 1.54 \times \sqrt{\left|\psi_{2,1} \right|} [mWb]$.

The evolution of the resonant helical magnetic flux following the application of a \SI{1}{ms} $n=1$ broad-$m$ pulse is shown in Fig. \ref{fig:psi_evo}. Figure \ref{fig:psi_evo}A only shows the evolution of the $2/1$ flux for clarity, whereas Fig. \ref{fig:psi_evo}B shows the evolution of multiple low-order resonant fluxes.

Focusing first on the $2/1$ mode, the evolution of this mode shows three distinct phases of growth. First, the amplitude of the helical resonant flux grows and decays with the application of the external magnetic pulse. The amplitude of the $2/1$ mode continues to decay following the termination of the pulse until about \SI{1.5}{ms}. At \SI{1.5}{ms} the $2/1$ mode transitions to a slow growth phase which persists until roughly \SI{8}{ms}. At \SI{8}{ms} the mode transitions into a faster robust growth phase.

This faster robust growth phase is consistent with a pressure-gradient-driven NTM. The linear growth of the helical flux is indicative of an NTM \cite{QuCallen1985}. (If the growth was classically driven then the helical flux should grow quadratically in time \cite{Rutherford1973}). The increased linear growth during the fast growth phase is described by a modified Rutherford equation analysis using the parameters for the initial reconstruction, as shown in Fig. \ref{fig:psi_evo}A. During the fast growth phase the $2/1$ grows linearly at a rate of approximately \SI{3.4}{Wb/s}. The MRE analysis using the simulated parameters predicts a growth of \SI{3.3}{Wb/s} \cite{Callen21}. This good agreement between the MRE analysis and simulation during the robust growth phase indicates that this growth is standard NTM growth.

The full dynamics involves multiple higher-order modes. The application of the external magnetic pulse also induces a large $3/1$ and $4/1$ (not shown) response in addition to the $2/1$. These helicities are resonant on surfaces located near the computational boundary where the MP is applied, and the close proximity of the surfaces to the wall limits the plasma screening of these helicities. The amplitude of the $3/1$ and $4/1$ rapidly grow with the rise of the applied MP pulse. The $4/1$ mode also decays with the decay of the pulse; however, the $3/1$ mode has a prolonged decay lasting several milliseconds. Following this initial decay, both the $3/1$ and $4/1$ resonant helical fluxes remain small through the remainder of the simulation.

\begin{figure}[!h]
\centering
\includegraphics[width=0.5\textwidth]{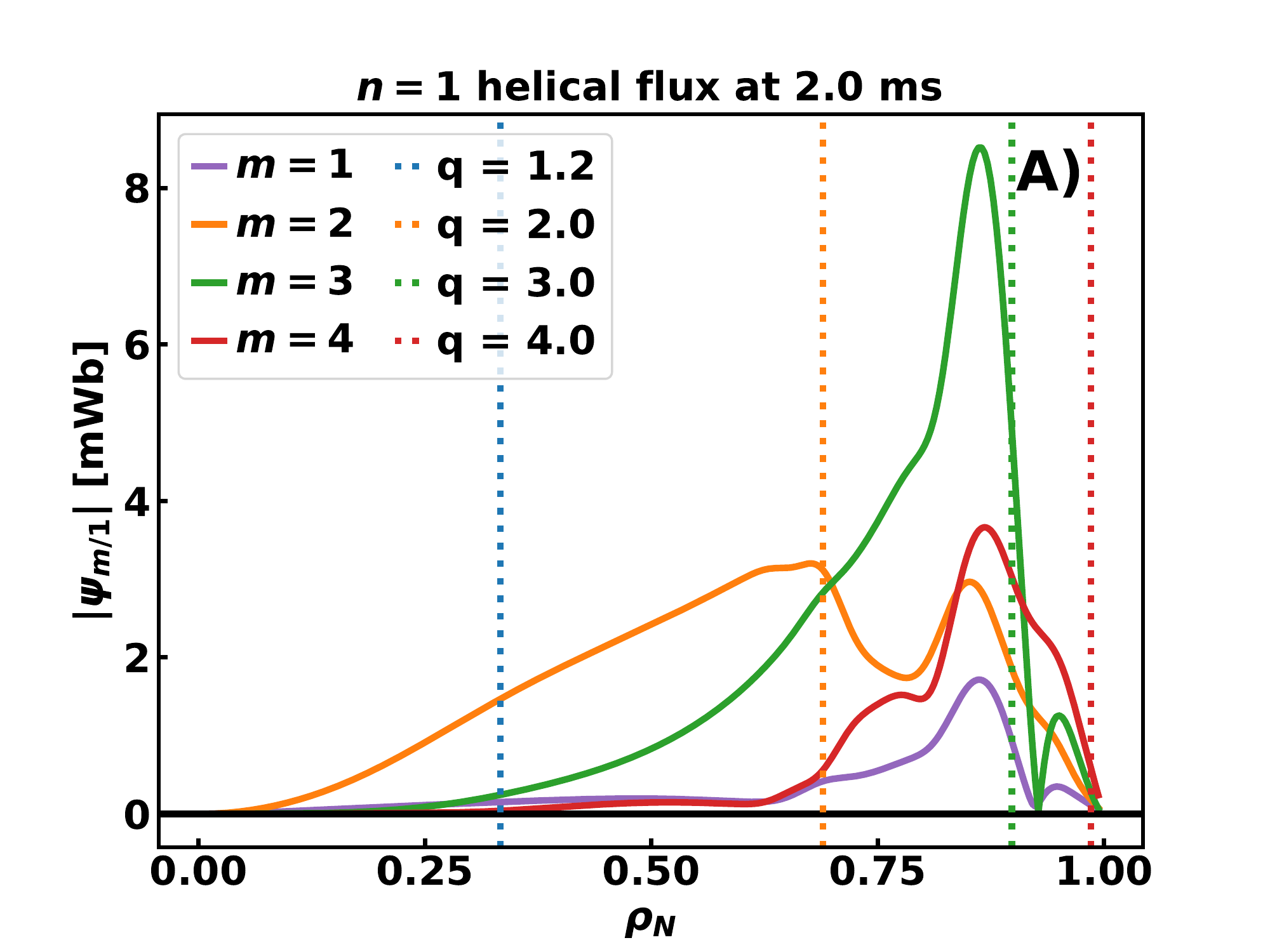}
\newline
\includegraphics[width=0.5\textwidth]{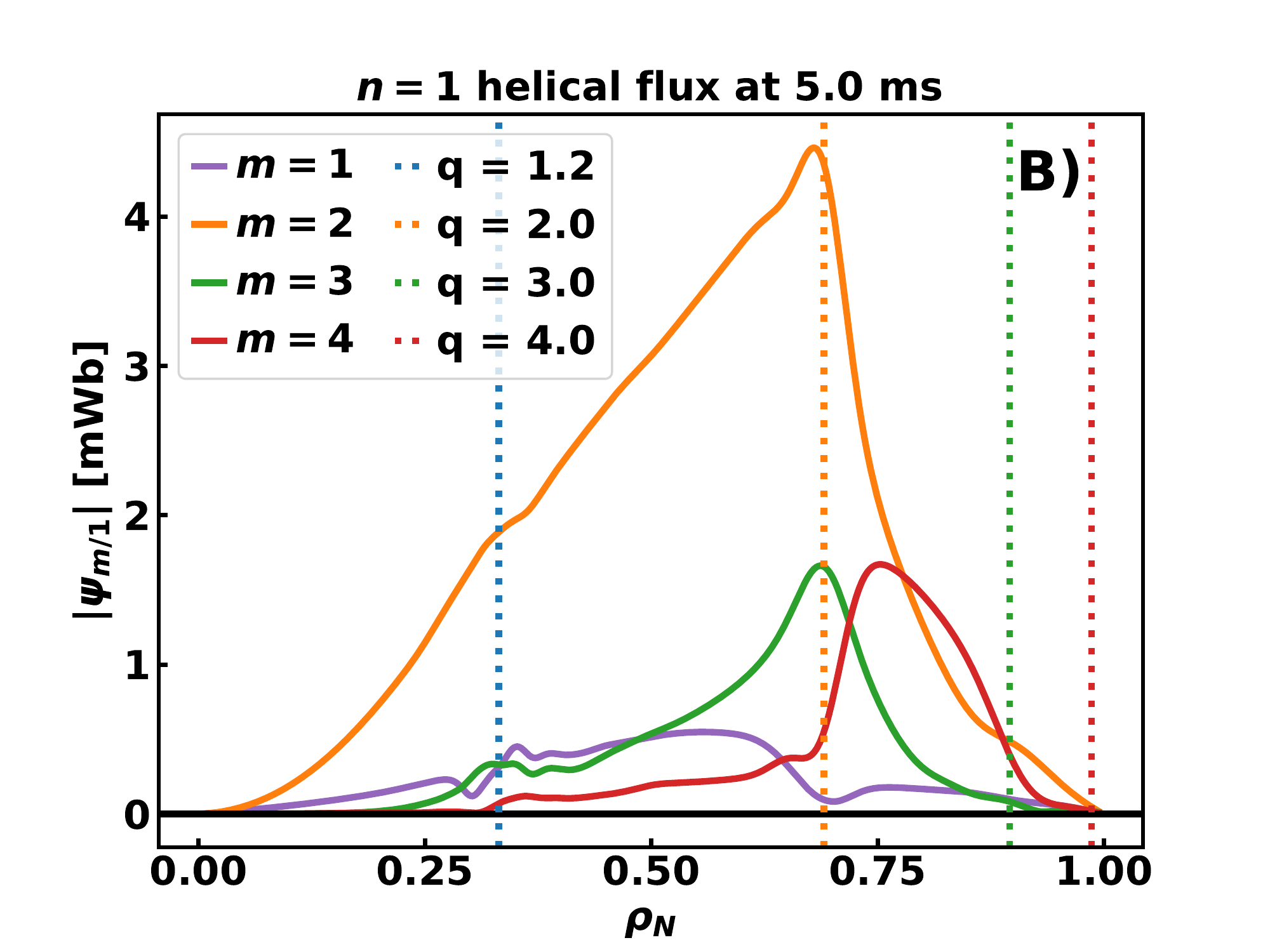}
\caption{The n=1 helical flux is plotted as a function of $\rho_N$ at A) \SI{2.0}{ms} and B) \SI{5.0}{ms}. The $3/1$ helical flux is dominant at \SI{2.0}{ms} and peaks inside of the $q=3$ surface. The nonzero $3/1$ flux at $q=3$ is indicative of tearing. The perturbation to the $1/1$, $2/1$, and $4/1$ fluxes near the $q=3$ surface suggests strong poloidal-mode coupling. The $2/1$ helical flux is dominant at \SI{5.0}{ms}, and the nonzero $2/1$ flux at $q=2$ is indicative of tearing. Note figures A) and B) use different vertical scales.}
\label{fig:radial_flux}
\end{figure}

Figure \ref{fig:radial_flux} shows the radial structure of the poloidal harmonics of the $n=1$ mode at \SI{2.0}{ms} and \SI{5.0}{ms} as a function of the normalized toroidal flux $\rho_N$. The time \SI{2.0}{ms} in Fig. \ref{fig:radial_flux}A follows the termination of the magnetic perturbation, and it is representative of conditions at the beginning of the $2/1$ slow growth phase. At this time the $3/1$ helical flux is slowly decaying but is still the dominant component. The $3/1$ helical flux peaks just inside the $q=3$ surface. The finite value of the resonant helical flux at the $q=3$ surface indicates tearing. Due to the poloidal-mode coupling the $4/1$, $2/1$, and $1/1$ helical fluxes also peak in this vicinity. The $2/1$ flux also has a local maximum near the $q=2$ surface, suggesting that there is strong coupling between the $2/1$ and $3/1$ modes.

The helical flux profile at \SI{5}{ms}, Fig. \ref{fig:radial_flux}B, is characteristic of the $n=1$ flux during the last half of the slow growth phase and into the robust growth phase. Here the $2/1$ component is the dominant $n=1$ component, and the finite resonant helical flux indicates tearing. Due to poloidal-mode coupling, there are $1/1$, $3/1$, and $4/1$ sidebands. The resonant $3/1$ and $4/1$ fluxes are small at $q=3$ and $q=4$, respectively. The small perturbations to the helical flux profile around $\rho_N = 0.35$ occur near the $q=1.2$ surface, and result from the nonlinear transfer of energy into the $n=1$ from the interaction between the $n=5$ and $n=6$ modes (see the discussion in Section \ref{sec:analysis}).

As seen in Fig. \ref{fig:psi_evo}B, high-$n$ core modes begin to grow in a sequence starting around \SI{4}{ms}. The $6/5$ mode grows out of the noise first, followed shortly thereafter by the $5/4$ and $4/3$ modes. The $3/2$ mode begins to grow once the $4/3$ mode reaches large amplitude; once the $3/2$ mode reaches large amplitude the $2/1$ mode transitions from the slow growth phase to the robust growth phase. Simultaneously, as modes in this sequence grow to large amplitude the previous mode sequence peaks in amplitude and then decays. For example, the $4/3$ mode peaks when the $3/2$ mode grows to large amplitude.

\begin{figure*}[t]
\centering
\subfloat{\includegraphics[width=0.5\textwidth]{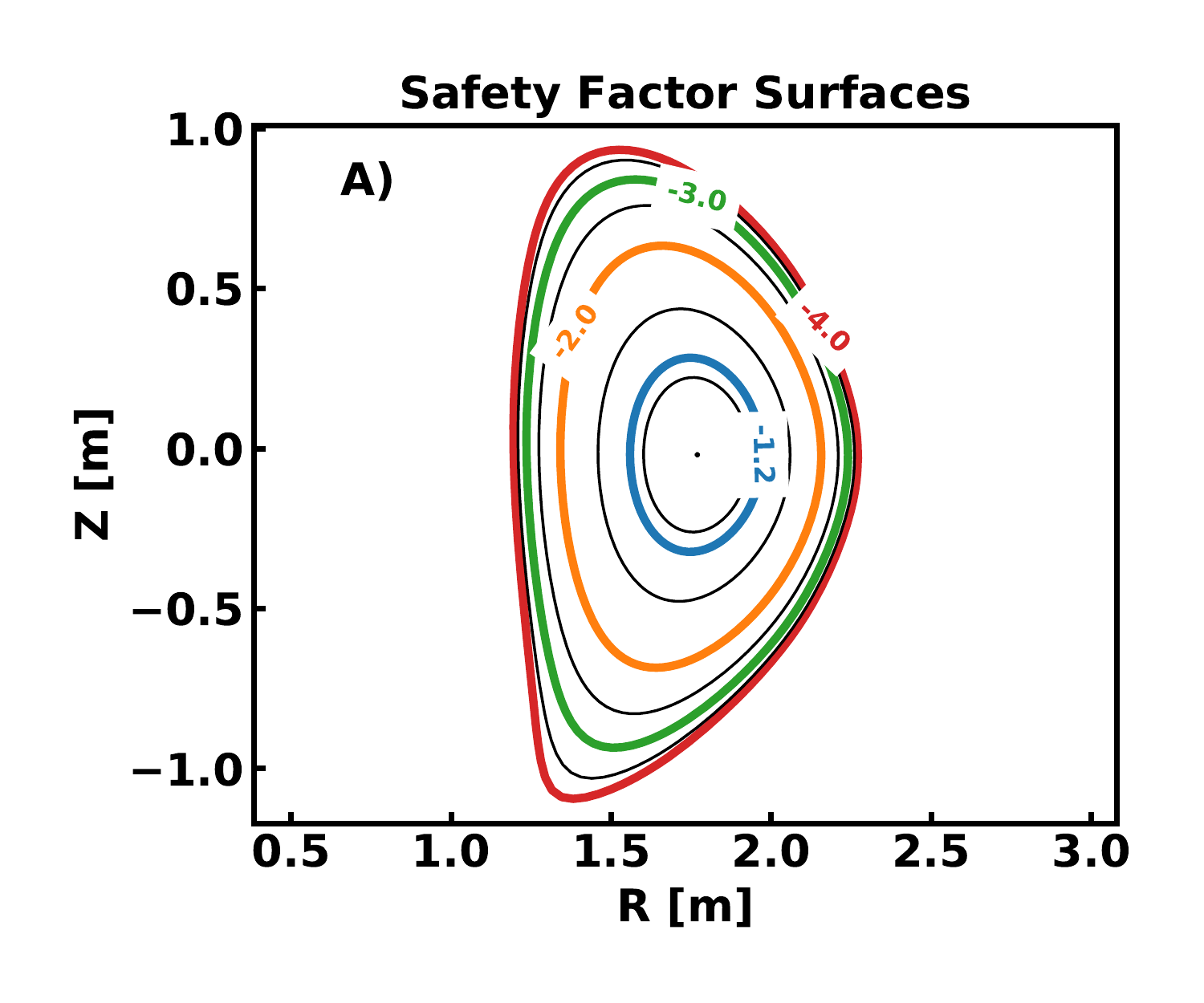}}%
\subfloat{\includegraphics[width=0.5\textwidth]{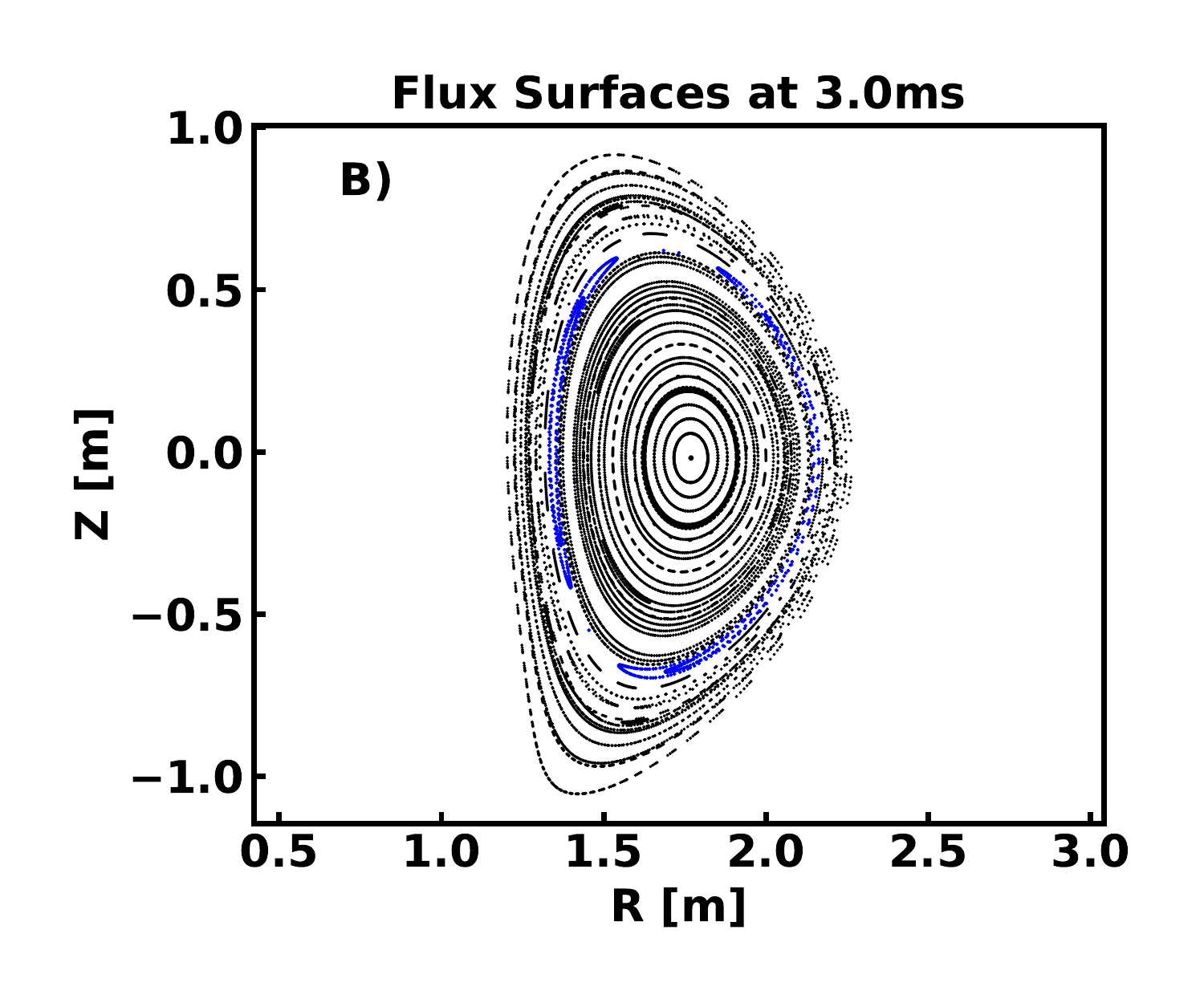}}%
\newline
\centering
\subfloat{\includegraphics[width=0.5\textwidth]{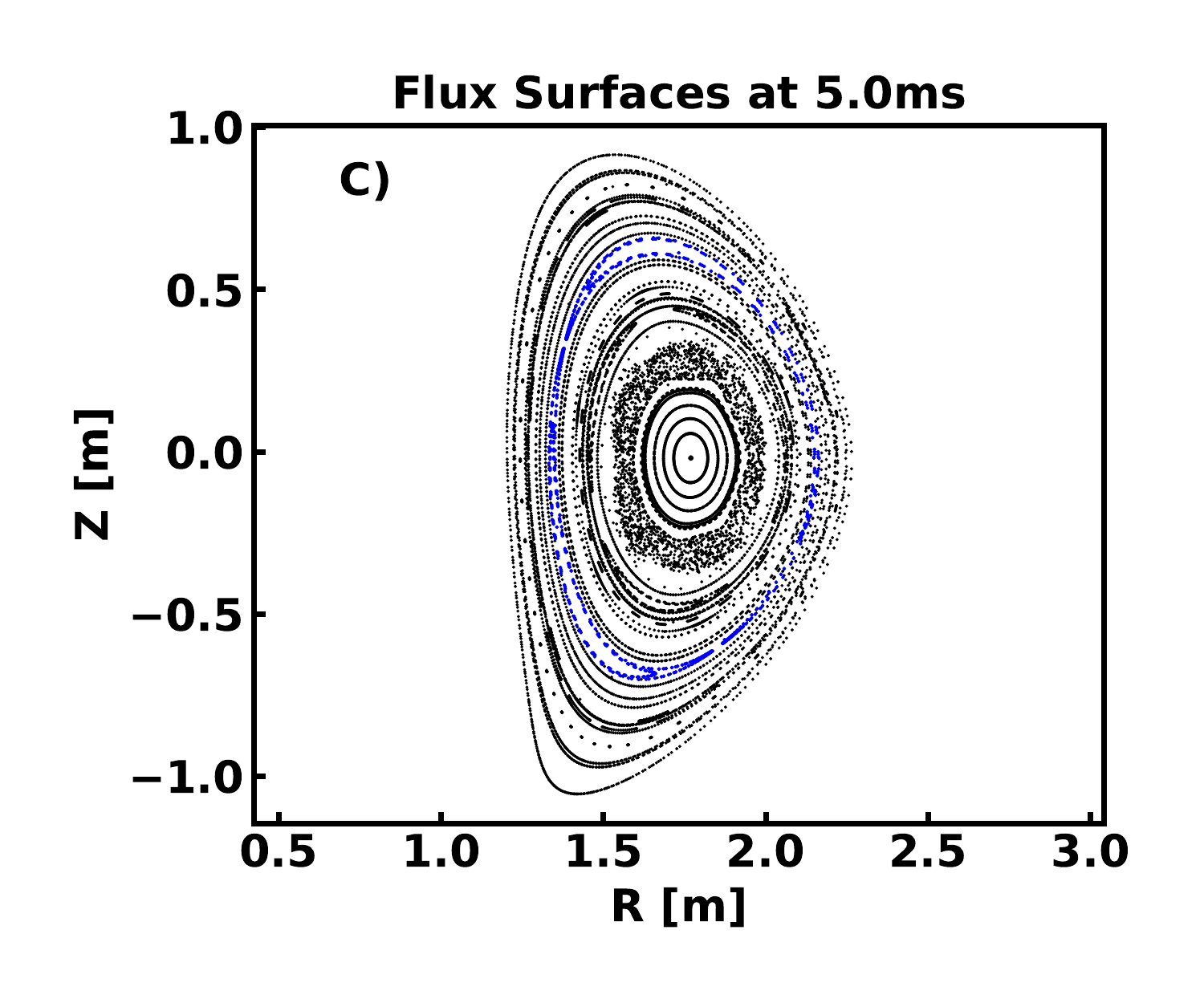}}%
\subfloat{\includegraphics[width=0.5\textwidth]{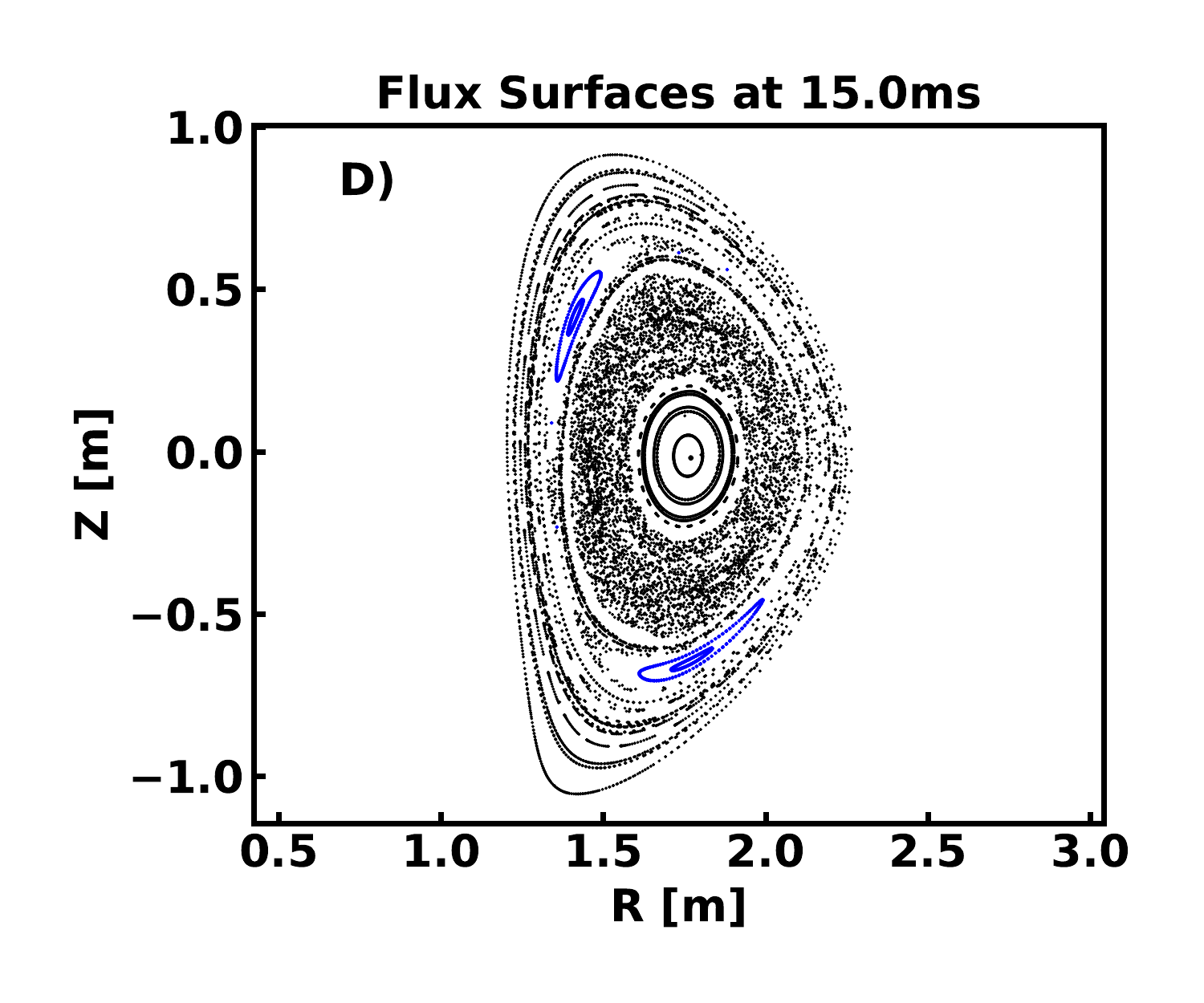}}%
\centering
\caption{The evolution of the magnetic topology is illustrated through a series of Poincare plots. A) The location of several low-order rational surfaces is shown for reference. B) At \SI{3}{ms} most of the surfaces are closed, and a thin $2/1$ island is indicated in blue. C) At \SI{5}{ms} the growth of high-order core modes degrades the core surfaces. D) At \SI{15}{ms} the surfaces inside $q=2$ are stochastic and a large $2/1$ is present.}
\label{fig:flux_surfaces}
\end{figure*}

The growth of the multiple modes creates islands that overlap and generates regions of stochastic magnetic field. Figure \ref{fig:flux_surfaces} illustrates the evolution of the magnetic topology throughout the simulation. Figure \ref{fig:flux_surfaces}A shows the equilibrium location of low-order rational surfaces for reference. Figure \ref{fig:flux_surfaces}B shows the magnetic topology at \SI{3}{ms}. This is during the slow growth phase of the $2/1$, after the $3/1$ mode has decayed to small amplitude and prior to the growth of the high-$n$ core modes. Here most surfaces are closed, and there is a small 2-\SI{3}{cm} $2/1$ island (indicated in blue). Figure \ref{fig:flux_surfaces}C shows the flux surfaces at \SI{5}{ms}; here the $2/1$ mode is slowly growing and several high-$n$ cores modes have grown to large amplitudes. The growth of these core modes creates a stochastic region in the core which spreads radially outwards as more core modes grow. The flux surfaces at \SI{15}{ms} are shown in Fig. \ref{fig:flux_surfaces}D. This time is characteristic of the robust $2/1$ growth phase. Here the $2/1$ island width has grown to \SI{5}{cm}, and the core stochastic region has spread radially outwards towards the $2/1$ island, but surfaces outside the $2/1$ are closed.

\begin{figure}[h]
\centering
\includegraphics[width=0.5\textwidth]{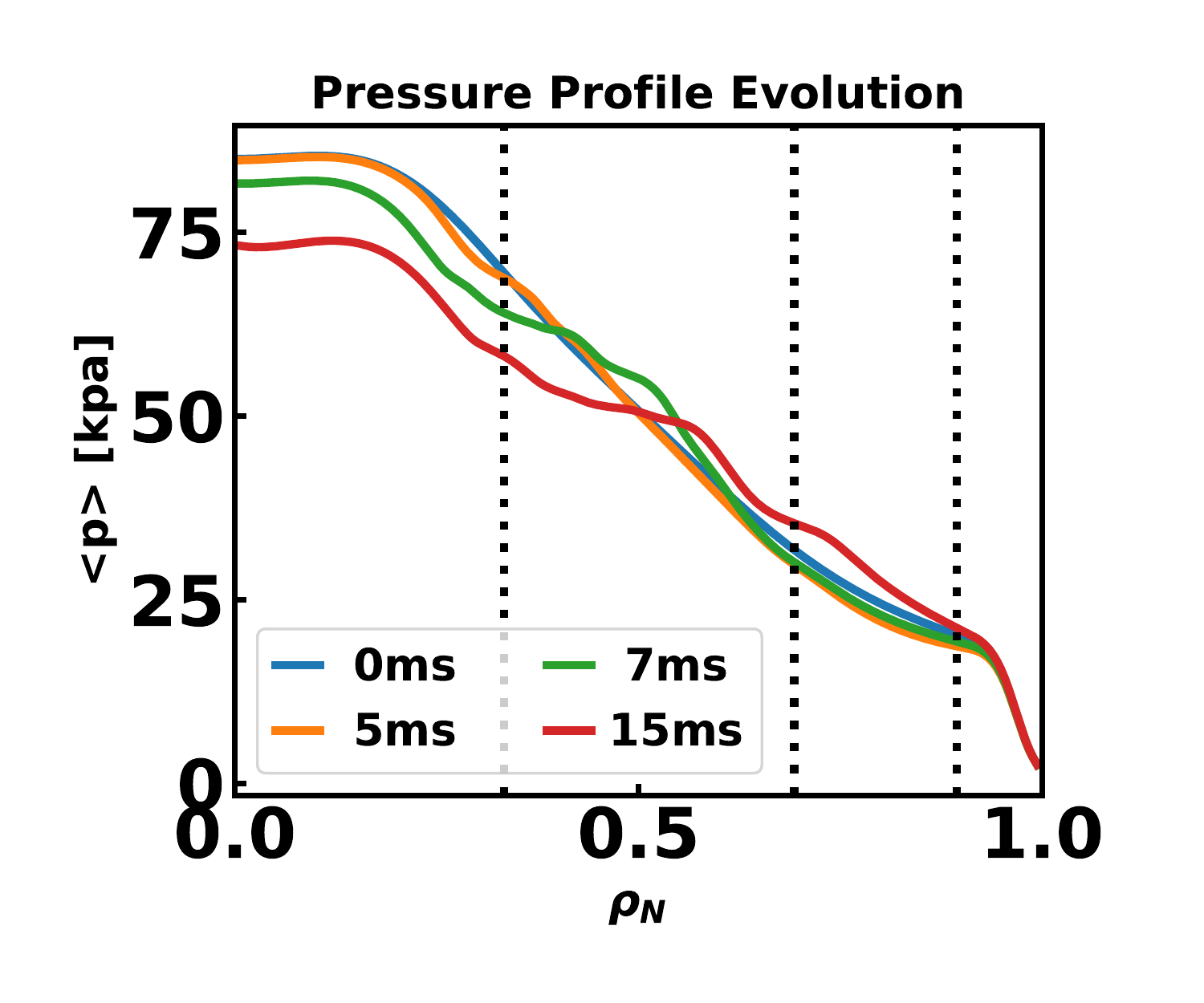}
\caption{Evolution of the flux surface averaged pressure profile. From left to right the vertical dotted lines indicate the location of the equilibrium $q=1.2$, $q=2$, and $q=3$ surfaces.}
\label{fig:fsap}
\end{figure}

Flattening of the pressure profile in the core occurs as a result of the stochastic flux region with anisotropic thermal conduction. Figure \ref{fig:fsap} shows the evolution of the flux surface averaged pressure profile. The pressure profile outside of the core stochastic region is relatively unperturbed, and the flattening of the pressure profile in the stochastic region increases the pressure gradient in the transition region from the stochastic region to closed surfaces. This quasi-linear change in the pressure gradient increases the instability drive. Its role in driving this sequence of core modes is explored in more detail in the next section. In particular, the next section address the question: ``Does the quasi-linear flattening or the nonlinear coupling between different modes drive the NTM growth?''

%% file: analysis.tex
\section{Power Transfer Analysis}\label{sec:analysis}

This section applies a power transfer Fourier mode analysis to understand the $2/1$ slow growth phase, lasting from approximately \SI{1.5}{ms} to \SI{8}{ms}, and contrasts it with the faster robust growth phase that persists from \SI{8}{ms} to the end of the simulation. The analysis calculates the power transfer between different toroidal Fourier modes, and it is a generalization of the cylindrical energy transfer analysis of Ho and Craddock \cite{Ho1991} to axisymmetric toroidal equilibria. A detailed derivation of the power transfer analysis is presented in Appendix \ref{apx:PowerTransfer}.

The analysis focuses on the energy transfer into and out of the non-axisymmetric, $n \neq 0$, toroidal Fourier modes. The volume integrated energy associated with the $n$-th non-axisymmetric Fourier mode is the sum of the kinetic and magnetic Fourier mode energies. (The axisymmetric $n=0$ energy also has a contribution due to the internal energy). The energy transfer between different toroidal modes results from the power associated with the kinetic and magnetic energies, and they are calculated from the magnetic induction equation and the momentum equation. The resulting power terms are grouped to highlight different physical effects. The following groupings are used: linear, quasi-linear, nonlinear, dissipative, Poynting flux, and advective. These groupings are defined in the appendix Eqs. (\ref{eqn:power_adv})-(\ref{eqn:power_pf}) and included in Table \ref{tab:powerTerms} for reference.

\begin{table}[h]
\renewcommand{\arraystretch}{2}
  \begin{tabular}{|c|l|}
    \hline
    \multirow{2}*{Linear} &
    $\left(\vec J_{eq} \times \vec B_n + \vec J_n \times \vec B_{eq}\right) \cdot \vec v_n^* + $ \\
    &$\left(\vec v_{eq} \times \vec B_n + \vec v_n \times \vec B_{eq}\right) \cdot \vec J_n^*$ - \\
    &$\nabla p_n\cdot \vec v_n^* + c.c.$ \\
    \hline
    \multirow{2}*{Quasi-linear} 
    &$ \left(\widetilde J_{0} \times \vec B_n + \vec J_n \times \widetilde B_{0}\right) \cdot \vec v_n^* +$\\
    &$\left(\widetilde v_{0} \times \vec B_n + \vec v_n \times \widetilde B_{0}\right) \cdot \vec J_n^* + c.c.$ \\
    \hline
    Nonlinear &  $\left(\widetilde J \times \widetilde B\right)_n \cdot \vec v_n^* + \left(\widetilde v \times \widetilde B\right)_n \cdot \vec J_n^*+ c.c.$\\
    \hline
    \multirow{2}*{Dissipative} & 
    $-\eta J_n^2 +\frac{1}{n_0e}\nabla \cdot \vec {\vec {\Pi}}_{e,n}\cdot \vec J_n^*+$\\
    &$\nabla \cdot \vec {\vec {\Pi}}_{i,n}\cdot \vec v_n^* + c.c.$
    \\
    \hline
    Poynting Flux & $-\nabla \cdot \frac{\vec E_n \times \vec B_n^*}{\mu_0} + c.c.$ \\
    \hline
    Advective & $-\rho_0 \left(\vec v \cdot \nabla \vec v\right)_n \cdot \vec v_n^*  + c.c.$ \\
    \hline
  \end{tabular}
\caption{The groupings of powers used in the power transfer analysis. Here $c.c.$ is the complex conjugate and $\widetilde f$ are perturbed quantities. The perturbed quantities in the nonlinear power exclude $n=0$ contributions to avoid double counting.}
\label{tab:powerTerms}
\end{table}

The linear, quasi-linear, and nonlinear powers quantify the ideal MHD drives that result from energy transfer between equilibrium, quasi-linear changes to the $n=0$ fields, and nonlinear multi-mode interactions. The Poynting flux is a conservative term that spatially redistributes the magnetic energy within a Fourier mode. The volume integral of the Poynting flux characterizes the energy injected into the control volume. The dissipative terms quantify the power associated with the ion viscosity, electron viscosity, and the electrical resistivity. The advective term quantifies the power transfer due to the center of mass advection, but is small here.

\begin{figure}[!h]
\centering
\includegraphics[width=0.5\textwidth]{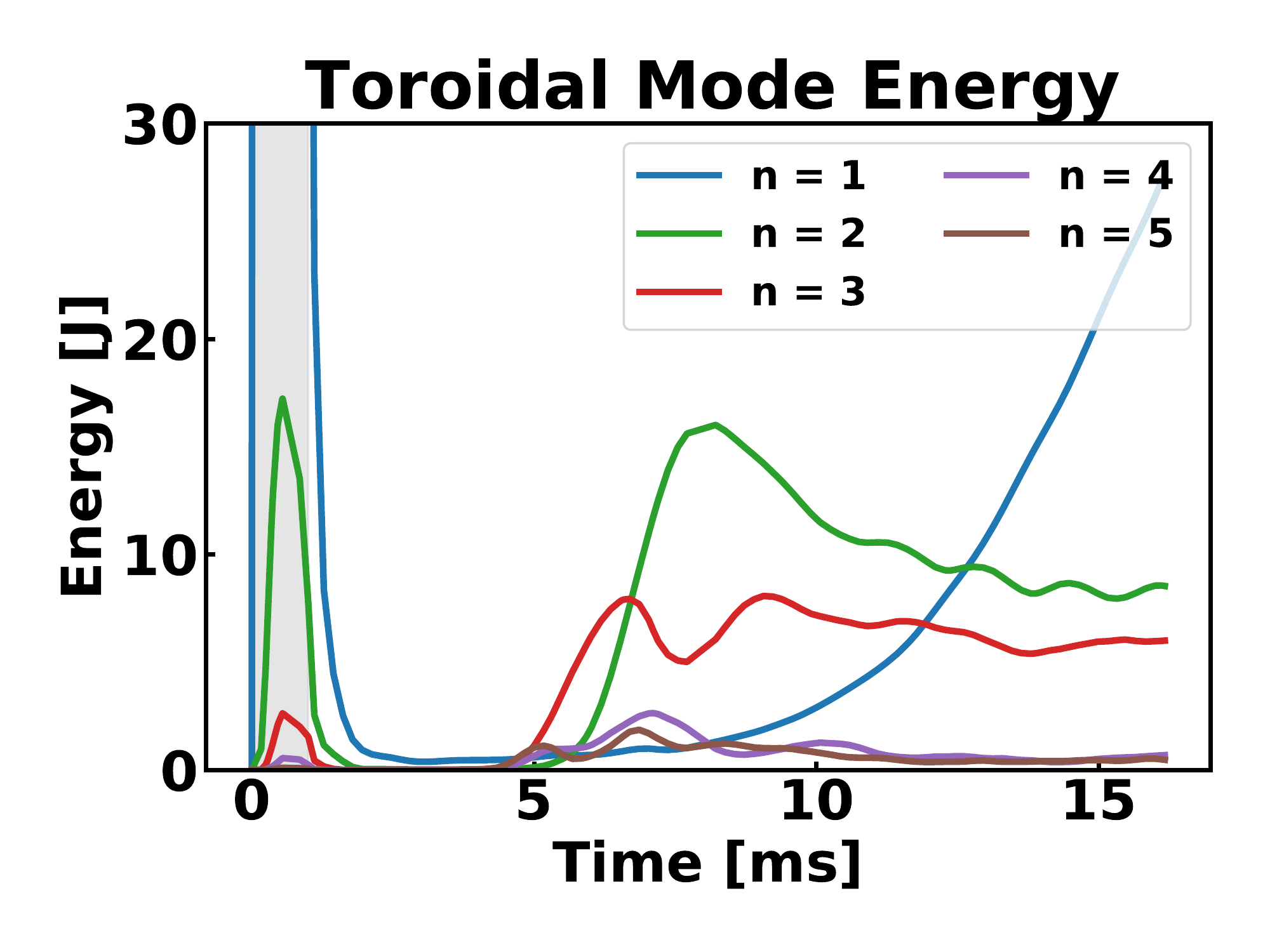}
\caption{Evolution of the total (kinetic + magnetic) toroidal Fourier mode energies for $n=1-5$. Similar to the evolution of the resonant helical flux in Fig. \ref{fig:psi_evo}, the $n=1$ energy grows in two phases following the \SI{1}{ms} MP.}
\label{fig:energyevo}
\end{figure}

One of the limitations of the power transfer analysis is that it provides no insight into the poloidal mode spectrum. Therefore it is useful to compare the evolution of the toroidal Fourier mode energy, shown in Fig. \ref{fig:energyevo}, with the evolution of the resonant helical fluxes, Fig. \ref{fig:psi_evo}B, in order to reorient the discussion. The two quantities provide a qualitatively similar view of the dynamics with some quantitative differences. The $n=1$ toroidal mode energy includes the energy in both $3/1$ and $2/1$ modes, and it experiences a slow growth phase that lasts from \SI{3}{ms} to \SI{8}{ms}. The onset of the slow growth in the $n=1$ energy is delayed relative to the onset of $2/1$ helical flux growth due to the decay of the $3/1$. In both the $n=1$ energy and the $2/1$ flux, the slow growth phase is followed by a fast robust growth. The transition to the fast growth phase is near instantaneous when considering the helical flux, but occurs over several ms in the magnetic energy. Similar to the helical flux, the magnetic energy also shows a sequence of high-$n$ modes growing up and saturating; however, the relative amplitude between different Fourier modes is different than the helical flux amplitudes. The energy associated with the $n>3$ Fourier modes is greatly reduced relative to the $n=1$, $2$, and $3$ Fourier modes. The higher-$n$ modes have smaller radial extents than the lower-$n$ modes, and their volume integrated energy is smaller as a result.

\begin{figure}[!h]
\centering
\includegraphics[width=0.5\textwidth]{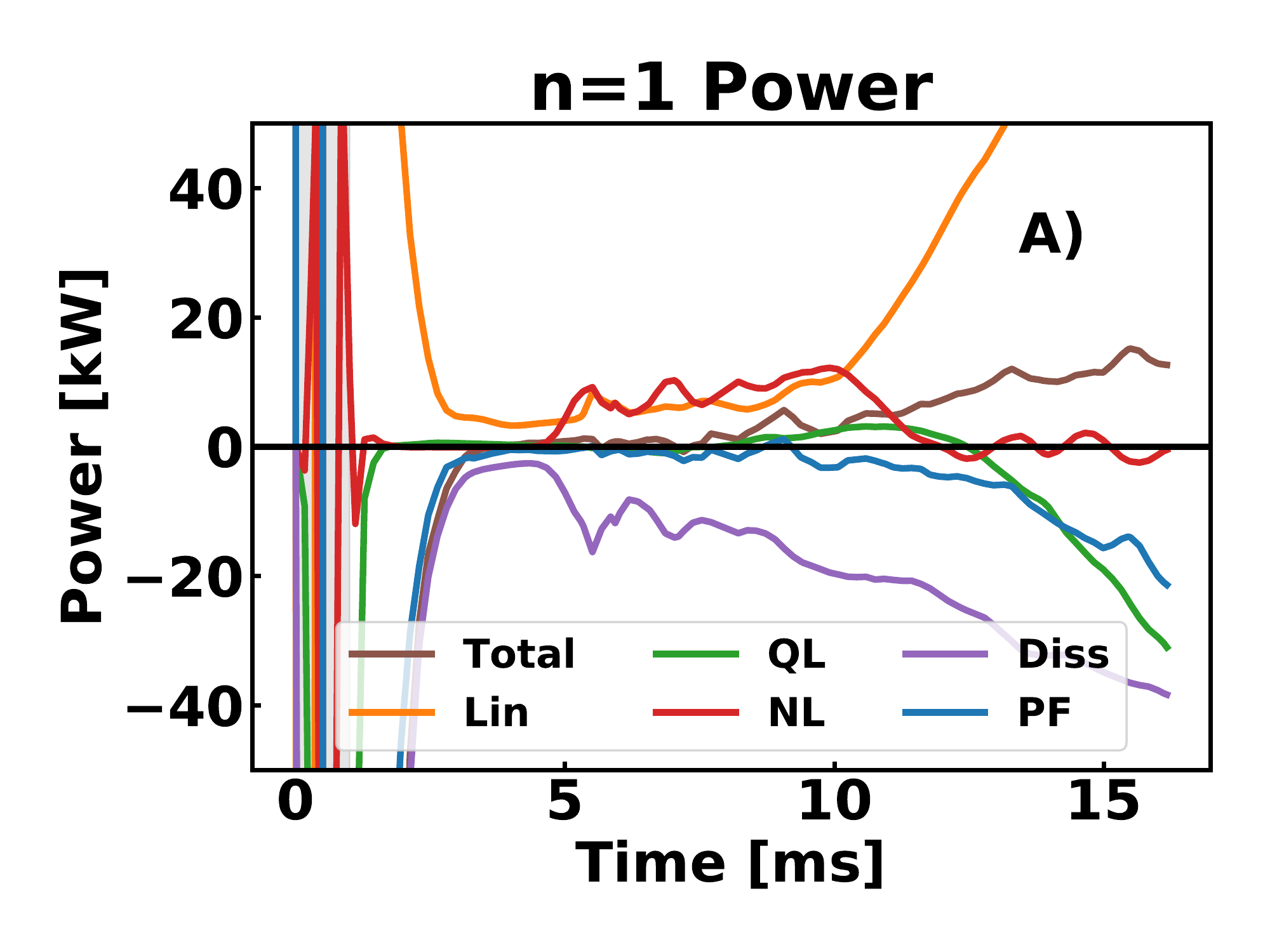}
\newline
\includegraphics[width=0.5\textwidth]{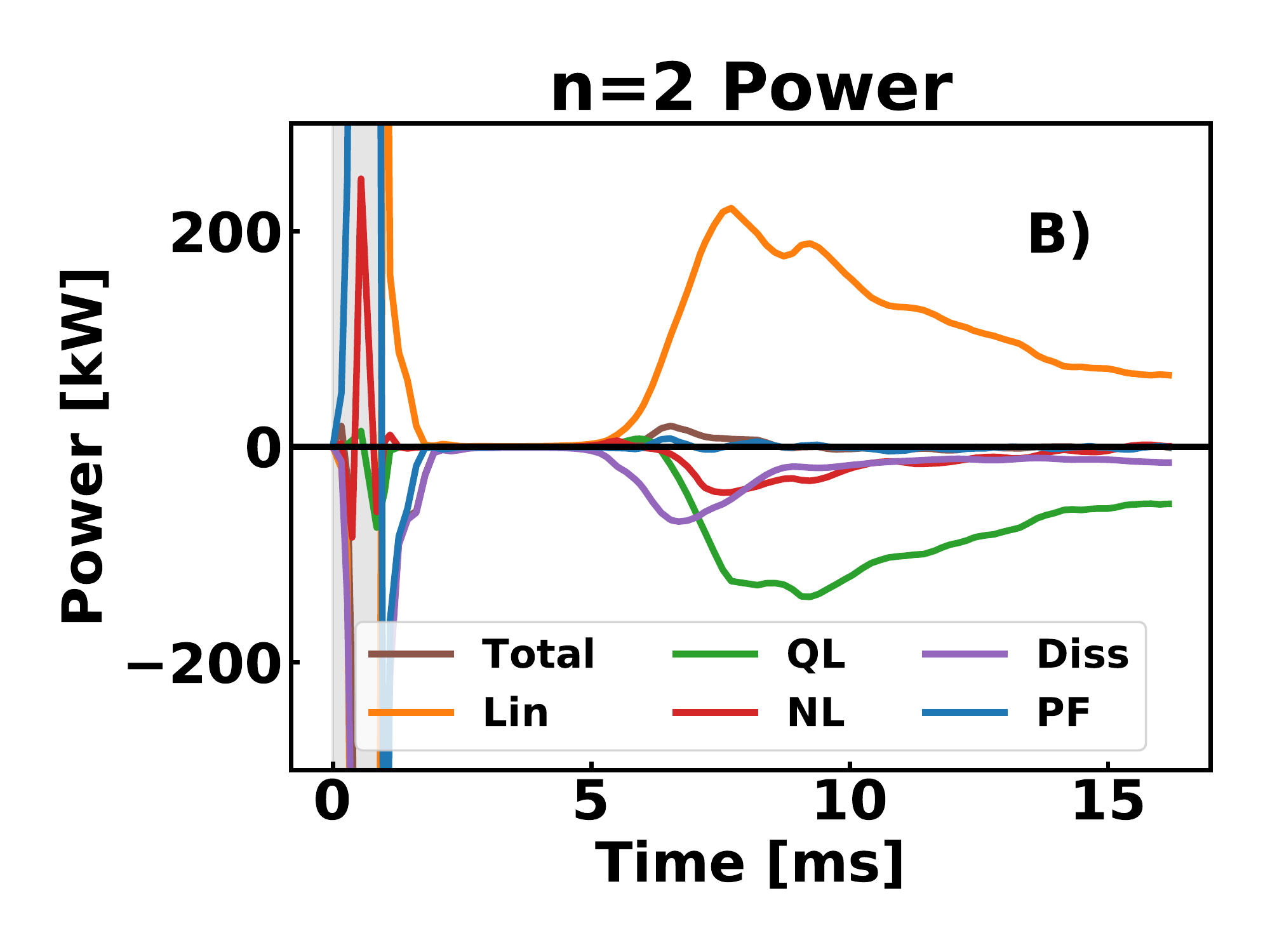}
\newline
\includegraphics[width=0.5\textwidth]{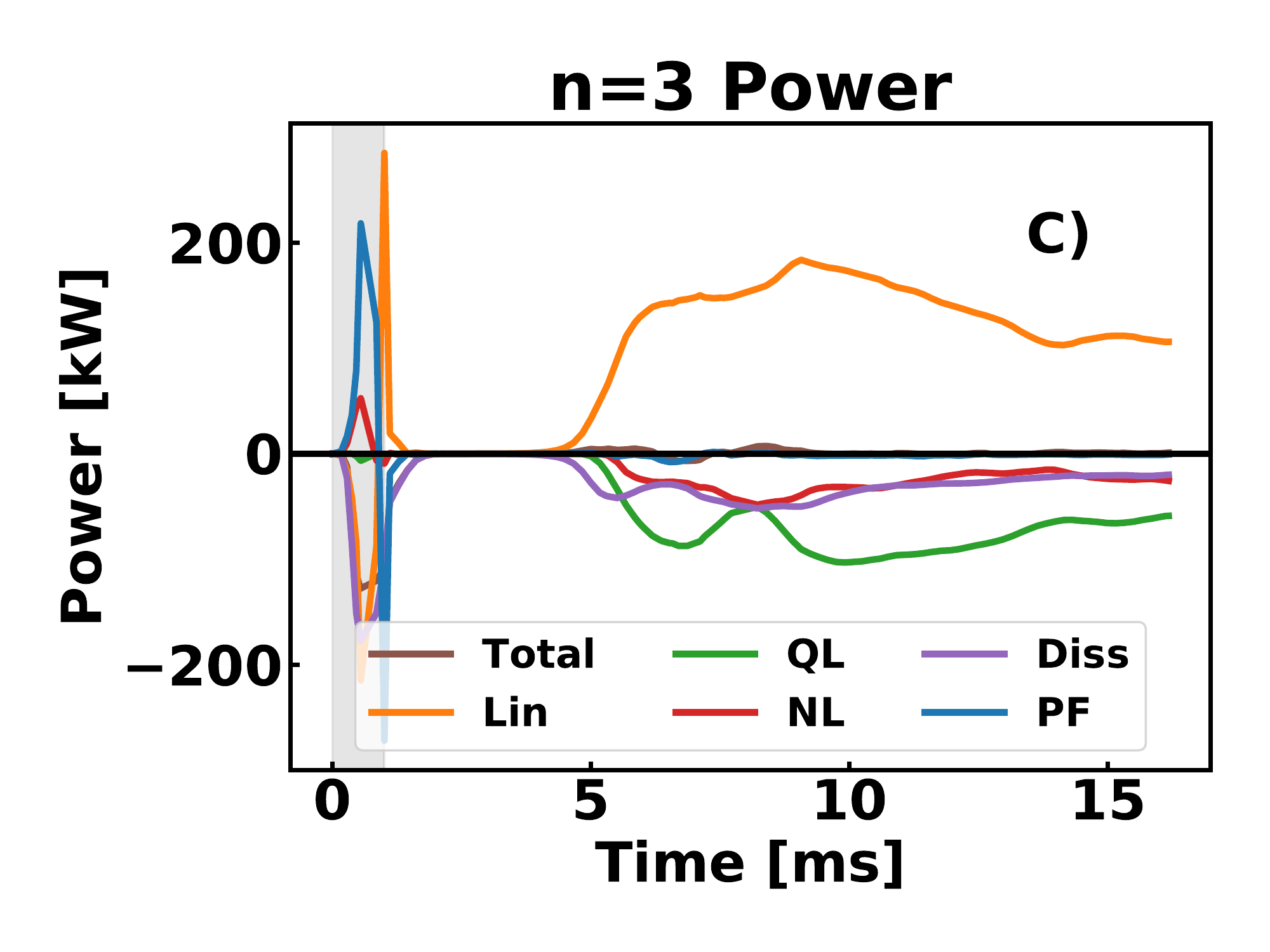}
\caption{Evolution of the volume integrated power for A) $n=1$, B) $n=2$ and C) $n=3$. The total power results from a small imbalance between the linear (lin), quasi-linear (QL), nonlinear (NL), dissipative (Diss), and Poynting flux (PF) powers. The nonlinear power is the significant power drive during the $n=1$ slow growth phase from \SI{5}{ms} to \SI{10}{ms}.}
\label{fig:total_power}
\end{figure}

The evolution of the volume integrated power is shown in Fig. \ref{fig:total_power} for $n=1$, $n=2$, and $n=3$. Starting with $n=2$ and $n=3$ (Figs. \ref{fig:total_power}B and \ref{fig:total_power}C), the total power into these Fourier modes results from a small imbalance between different terms. (The magnitude of the total power is much smaller than magnitude of the different terms). In both cases the linear terms are large and positive. This indicates that the power driving the growth of the $n=2$ and $n=3$ comes from the equilibrium fields. This is expected since most of the energy is in the equilibrium. The large linear power does not necessarily indicate linear instability. Linear \textsc{nimrod} simulations are stable, indicating that the $n=2$ and $n=3$ are linearly stable here.

During the initial growth of both $n=2$ and $n=3$ the linear power transfer is largely balanced by the dissipative power. This is characteristic of both classical and neoclassical tearing modes. The powers associated with the resistive and neoclassical electron stress are negative. These terms enable growth by enabling changes to the $2/1$ mode structure, which allows the mode to more effectively draw energy from the equilibrium. 

As the $n=2$ and $n=3$ Fourier modes grow to large energy the quasi-linear and nonlinear powers become significant and help balance the linear power. The growth of the quasi-linear power quantifies the stabilization due to the flattening of the axisymmetric current and pressure gradients in the vicinity of the dominant mode's rational surface. Following the initial $n=2$ growth, the $n=2$ quasi-linear power is the dominant term that balances the $n=2$ linear power from roughly \SI{8}{ms} onwards. The quasi-linear flattening is largely responsible for the saturation and subsequent decay of the $n=2$. While negative during this time, the nonlinear and dissipative powers are small in comparison, and they have a minor role in the saturation of the $n=2$.

In contrast, while the quasi-linear power is still the largest stabilizing term following the initial $n=3$ growth, the dissipative and nonlinear powers are  significant. The quasi-linear, nonlinear, and dissipative powers are all comparable around \SI{9}{ms}, and even late in the simulation the nonlinear and dissipative power combined represent 45\% of the power transfer out the mode.

The $n=1$ power transfer dynamics in Fig. \ref{fig:total_power}A are qualitatively different. From about \SI{3}{ms} to \SI{5}{ms}, the total power into the $n=1$ is approximately zero, and the total $n=1$ energy is relatively constant. During this time there is a small positive linear power that is balanced by the dissipative power. While the total power into the $n=1$ is approximately zero at this time, the $3/1$ resonant helical flux is decaying and the $2/1$ helical flux is slowly growing. We infer that during this time the growth of the $2/1$ is driven by power transfer from the $3/1$ to the $2/1$. However, the power transfer diagnostic cannot directly quantify this power transfer.

A positive nonlinear power grows up just prior to \SI{5}{ms}, and from 5 to \SI{10}{ms} the nonlinear power and linear powers contribute roughly equally to the power transfer into the $n=1$. The initial growth of this nonlinear power is coincident with initial growth of the higher-$n$ core modes. Around \SI{8}{ms}, when the $n=2$ energy is peaking, a positive quasi-linear power grows up and adds to $n=1$ power, but this power is smaller than the nonlinear power. The growth of the quasi-linear drive is consistent with the added drive due to the steepening of the pressure profile radially inwards of the $2/1$ surface. During the period from 5 to \SI{10}{ms} a small imbalance between the linear, nonlinear, quasi-linear and the negative dissipative power results in a small net positive power into the $n=1$, driving the slow $n=1$ growth. Starting around \SI{10}{ms} the linear power rapidly increases, the nonlinear power becomes small, the positive quasi-linear power peaks and then becomes negative, and the total power increases significantly. These changes occur as the $n=1$ transitions to the robust growth phase. 

\begin{figure}[!h]
\centering
\includegraphics[width=0.5\textwidth]{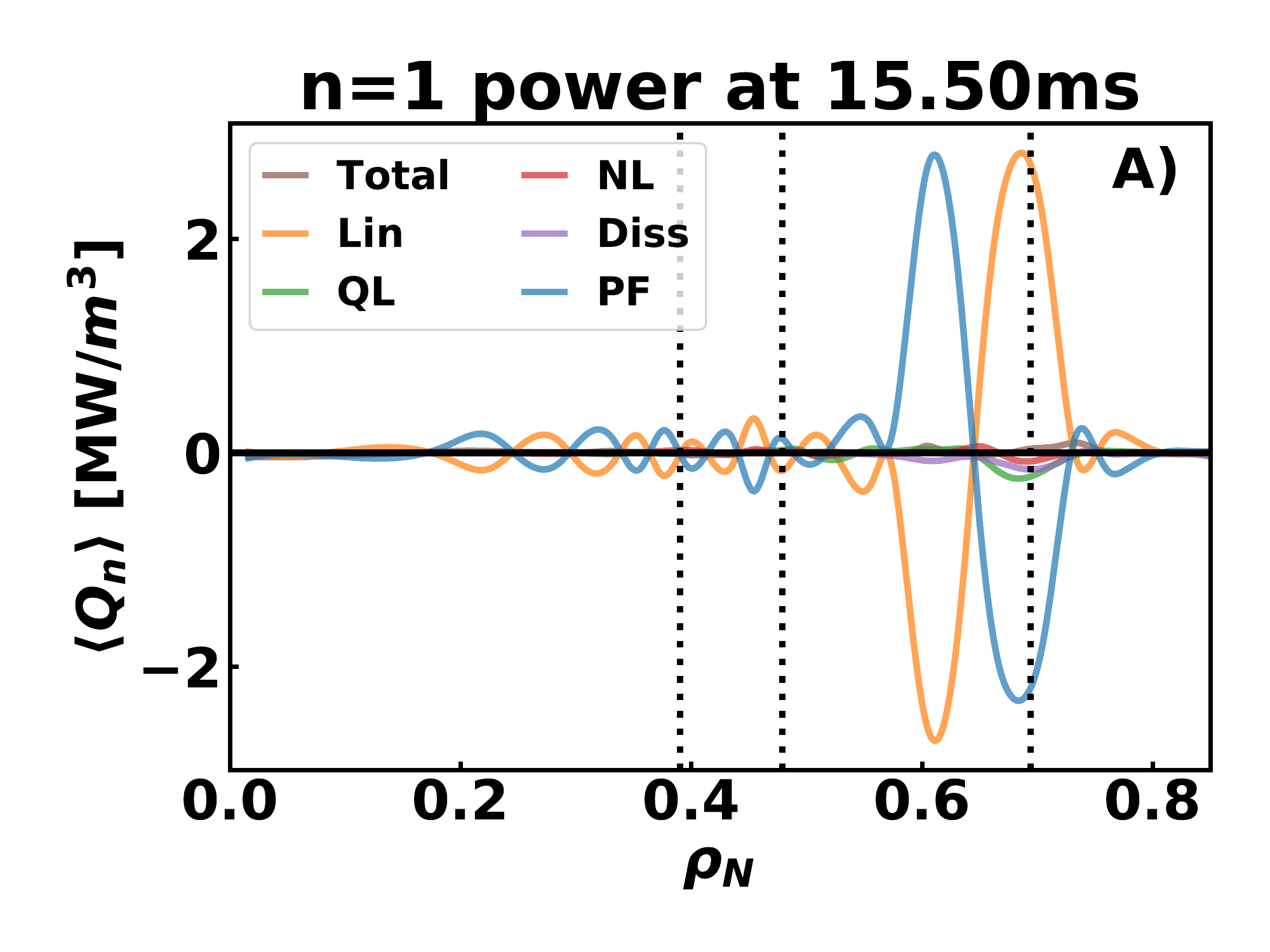}
\newline
\includegraphics[width=0.5\textwidth]{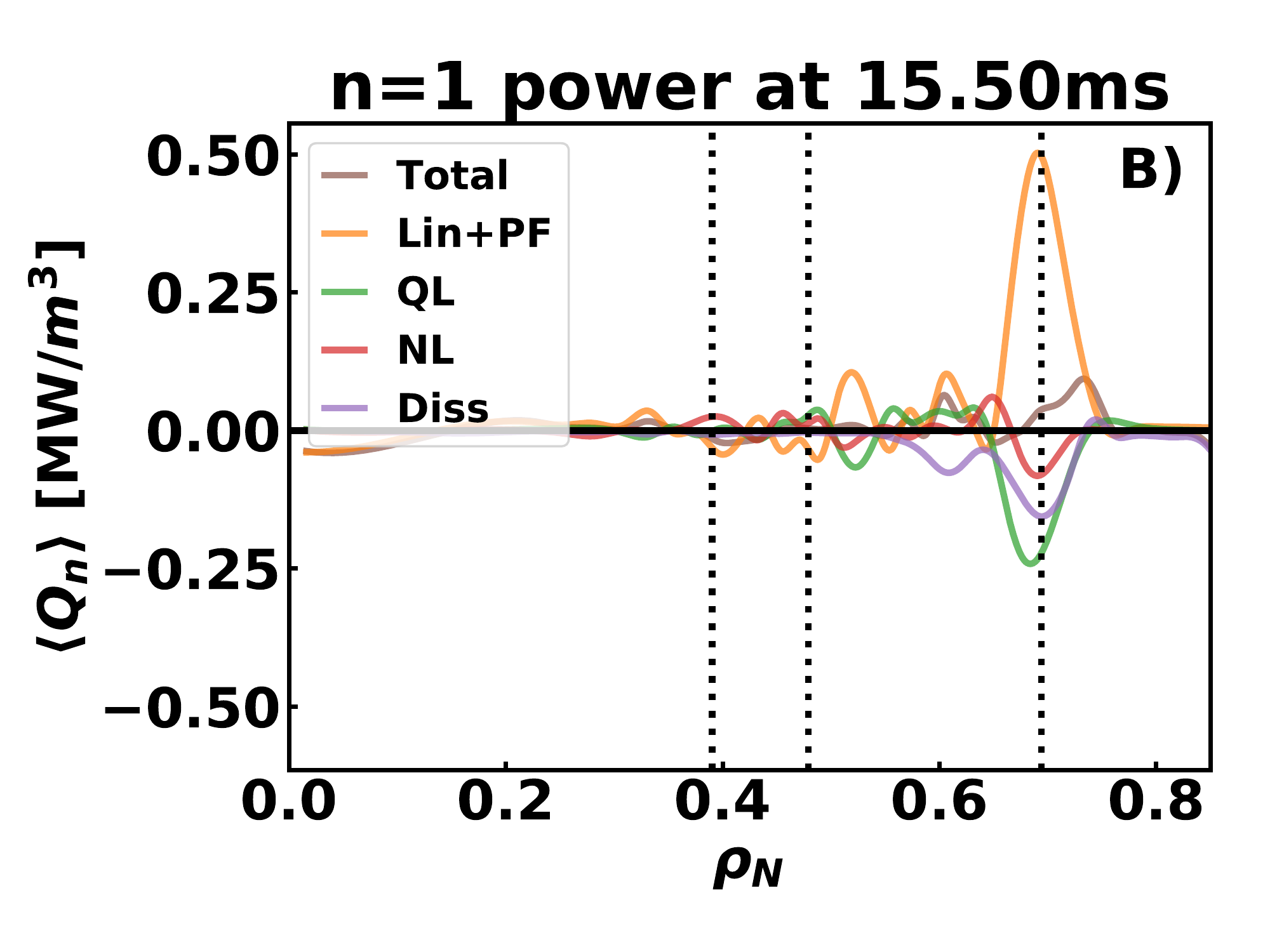}
\caption{FSA power at \SI{15.5}{ms}. A) The linear power (Lin) and Poynting flux (PF) powers largely cancel and are added together in B) for clarity. The resulting power into the $n=1$ is localized around the $q=2$ surface ($\rho_N=0.69$). From left to right the vertical dashed lines indicate the location of the $q=1.2$, $1.5$ and $2$ surfaces.}
\label{fig:fsa_power_15ms}
\end{figure}

The flux-surfaced averaged radial power deposition provides more insight into the dynamics. Three times are considered: \SI{5}{ms}, \SI{8.4}ms and \SI{15.5}ms. The power transfer into the $n=1$  at the last time \SI{15.5}{ms} is shown in Fig. \ref{fig:fsa_power_15ms}. The radial power deposition here is characteristic of the robust growth phase. As seen in Fig. \ref{fig:fsa_power_15ms}A, the two largest terms are the linear power and the Poynting flux. These terms are oscillatory and largely cancel. Near the $q=2$ surface located at $\rho_N = 0.69$, the linear power is positive and the Poynting flux power is negative. Radially inwards, at $\rho_N = 0.61$, the signs are reversed and here the Poynting flux power is positive and the linear power is negative. Physically, the $n=1$ is drawing energy from the equilibrium fields around the $q=2$ surface. The Poynting flux then transports most (about 80\%) of the injected energy inwards, where the linear powers transfers the energy back to the equilibrium. For clarity, the linear and Poynting flux powers are added together in Fig. \ref{fig:fsa_power_15ms}B, and the resulting sum demonstrates that the net power into the $n=1$ mode from these two terms is localized around the $q=2$ surface. This net power is largely balanced by the quasilinear and dissipative terms which are also strongly localized around the $q=2$ surface.

This localized power around a rational surface is the common signature for a island. In general, the sum of linear and Poynting flux powers will be positive in a narrow region around the rational surface. This power is largely balanced by a combination of the dissipative, quasi-linear, and nonlinear powers. When the island is small, the quasi-linear and nonlinear powers are generally small, and the dissipative power is the main term balancing the combined linear and Poynting flux power. The magnitude of the quasi-linear and nonlinear powers increases as the island grows.

\begin{figure}[!h]
\centering
\includegraphics[width=0.5\textwidth]{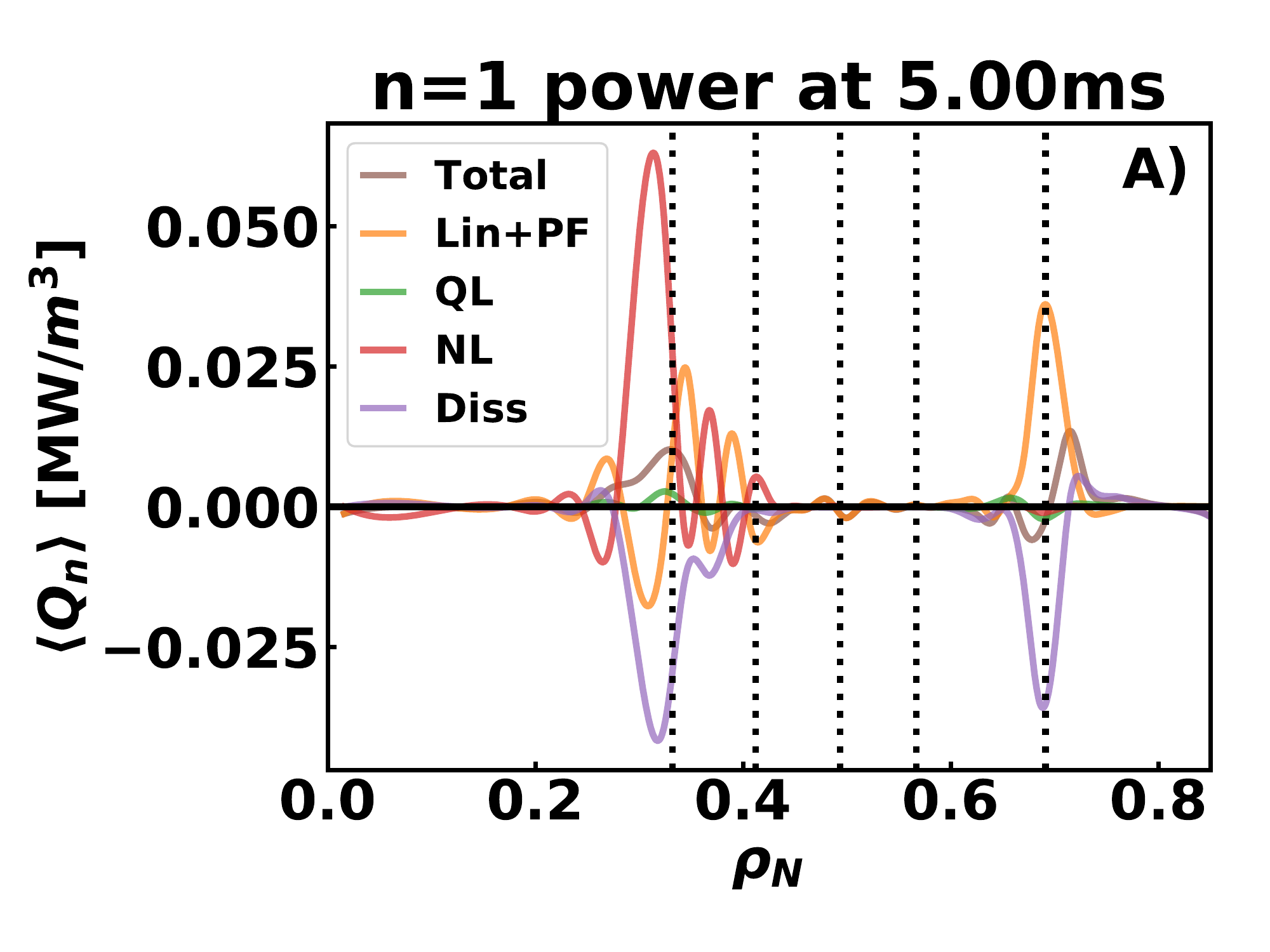}
\newline
\includegraphics[width=0.5\textwidth]{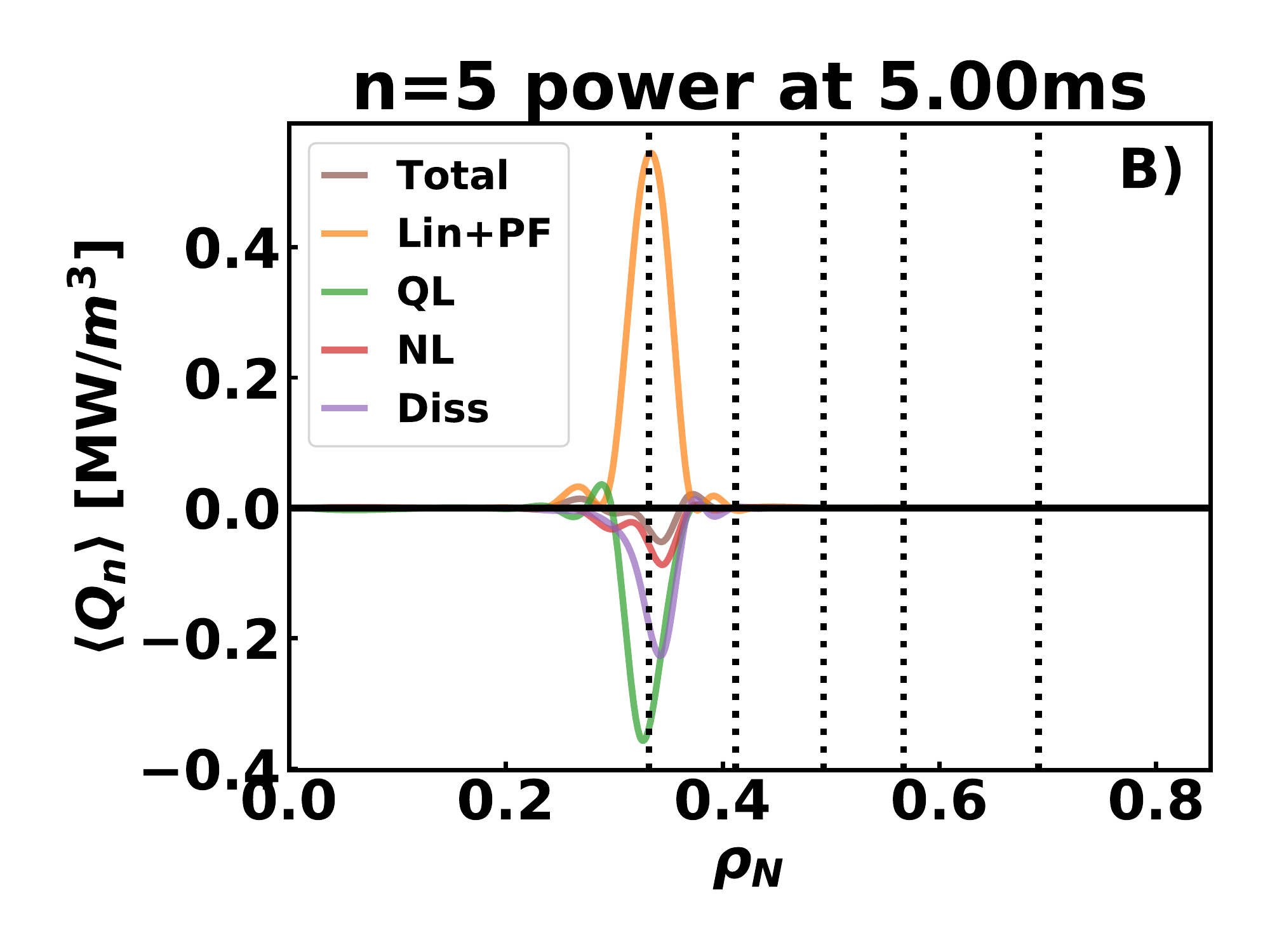}
\newline
\includegraphics[width=0.5\textwidth]{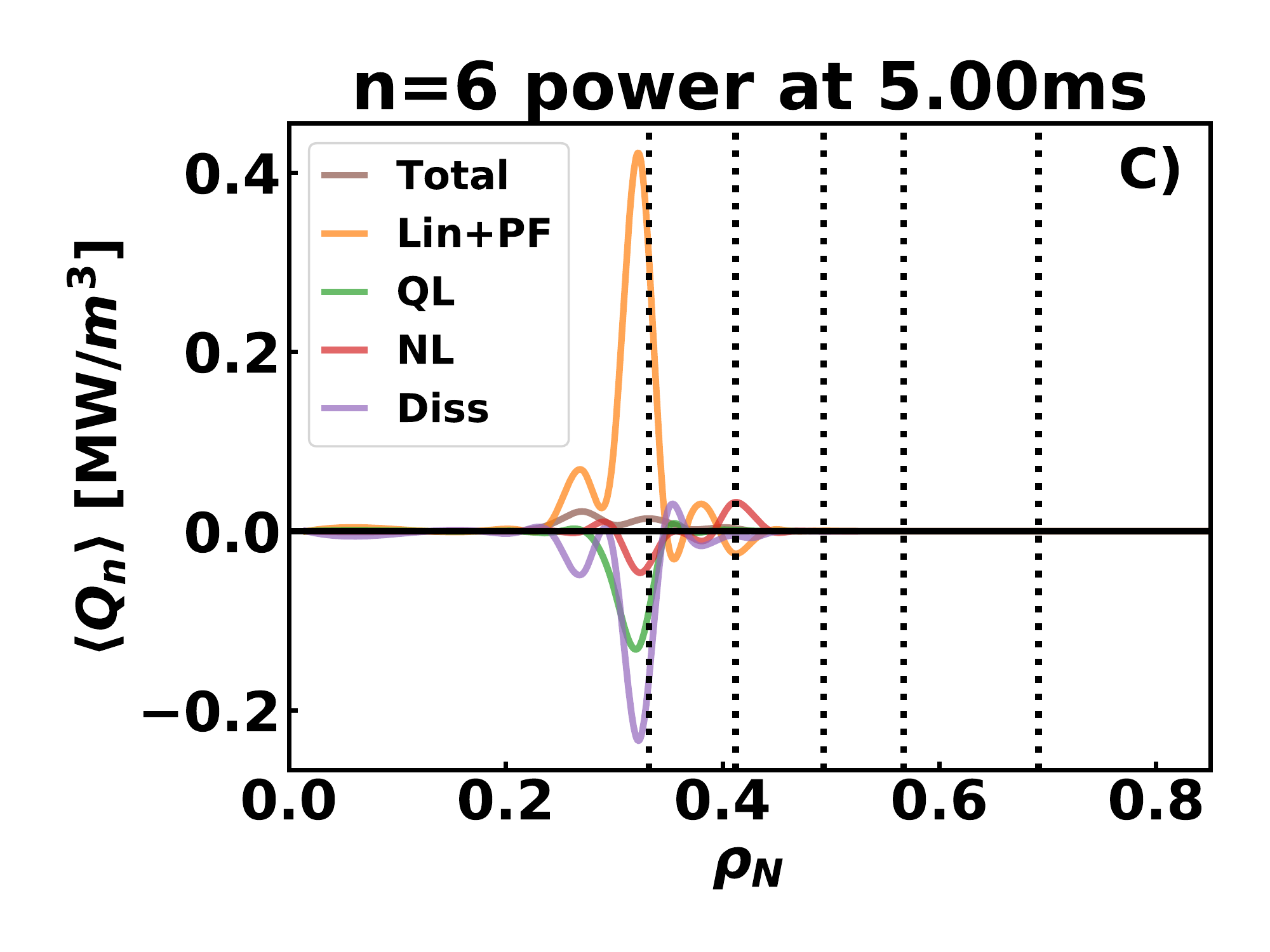}
\caption{FSA power at \SI{5.0}{ms}. A) The $n=1$ is localized around the two regions: $q \approx 1.2$ and $q=2$. The large nonlinear power at $q \approx 1.2$ results from the interaction between the 6/5 and 7/6. The B) $n=5$ and C) $n=6$ powers are localized around the $q=1.2$ and $q=1.17$ respectively.From left to right, the dotted vertical lines indicate $q=1.2, 1.33, 1.5, 1.67,\text{ and } 2.0$ surfaces.}
\label{fig:fsa_power_5ms}
\end{figure}

Next consider the time \SI{5}{ms}, which is the time during the slow growth phase when the nonlinear power transfer into the $n=1$ becomes significant, and the $n=5$ energy is near a local maximum. Figure \ref{fig:fsa_power_5ms}A shows the $n=1$ radial power distribution at this time. The linear and Poynting flux powers are added together, and their sum results in a net positive power into the $n=1$ in a region localized around $q=2$. In this region the injected energy is largely balanced by the dissipative power. However, there is also a large positive nonlinear power that is localized around $\rho_N = 0.31$ near $q=6/5$. The net power is positive in this region, but the nonlinear power is offset by the dissipative and combined linear plus Poynting flux power. As previously noted, this injected power in the vicinity of the $6/5$ surface has visible impact on the $n=1$ helical flux profile (Fig. \ref{fig:radial_flux}B). The $2/1$ component of the helical flux is flattened near the $6/5$ surface. The $1/1$, $3/1$, and $4/1$ helical fluxes are also distorted in this region.

The $n=5$ power at \SI{5}{ms} in Fig. \ref{fig:fsa_power_5ms}B is localized around $q=6/5$ and is characteristic of an island. This is consistent with the large resonant $6/5$ helical flux in Fig. \ref{fig:psi_evo}B at this time. The saturation of the $n=5$ energy is largely balanced by the quasilinear power, but there is a small negative nonlinear power. The $n=6$ power at this time in Fig. \ref{fig:fsa_power_5ms}C is localized around $q=7/6$, which is located inside of $q=6/5$. The $n=6$ combined linear and Poynting flux power is largely balanced by the dissipative power, but again there is a small negative nonlinear power. This suggests that the nonlinear 3-wave interaction between the $n=5$ and $n=6$ transfers energy from the $n=5$ and $n=6$ to the $n=1$. This argument is further supported by the fact that the $n=1$, $5$, and $6$ nonlinear powers are comparable in amplitude in this region. The $n=5$ nonlinear power has a maximum amplitude of \SI{-0.086}{MW/m^3}, the $n=6$ nonlinear power has a maximum amplitude of \SI{-0.046}{MW/m^3} and the nonlinear power into the $n=1$ has a maximum of \SI{0.063}{MW/m^3}. The fact that the powers don't exactly sum to zero is explained by noting that power is also nonlinearly injected into the $n=2$ and $n=4$ modes in this region.

\begin{figure}[h]
\centering
\includegraphics[width=0.5\textwidth]{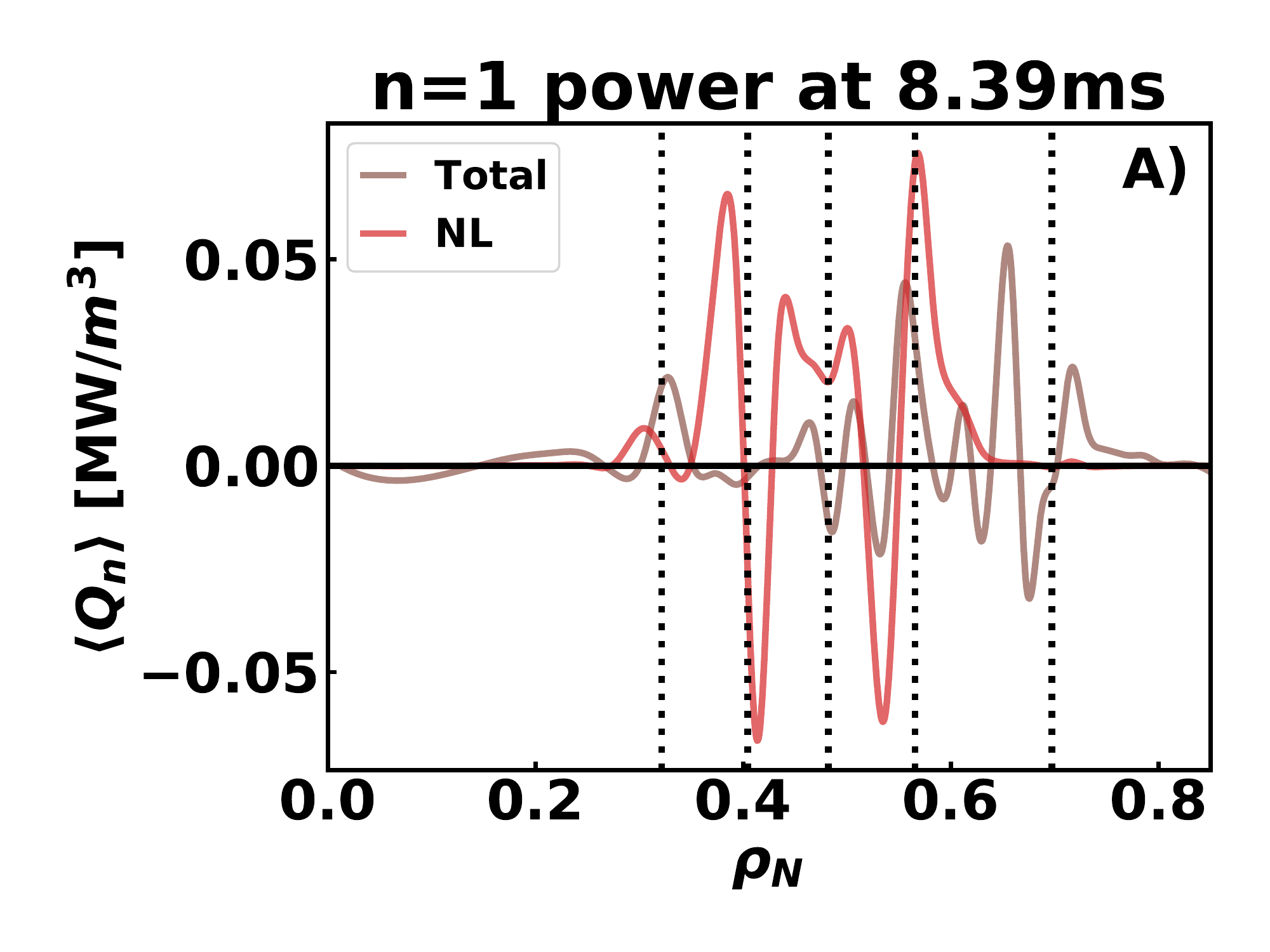}
\newline
\includegraphics[width=0.5\textwidth]{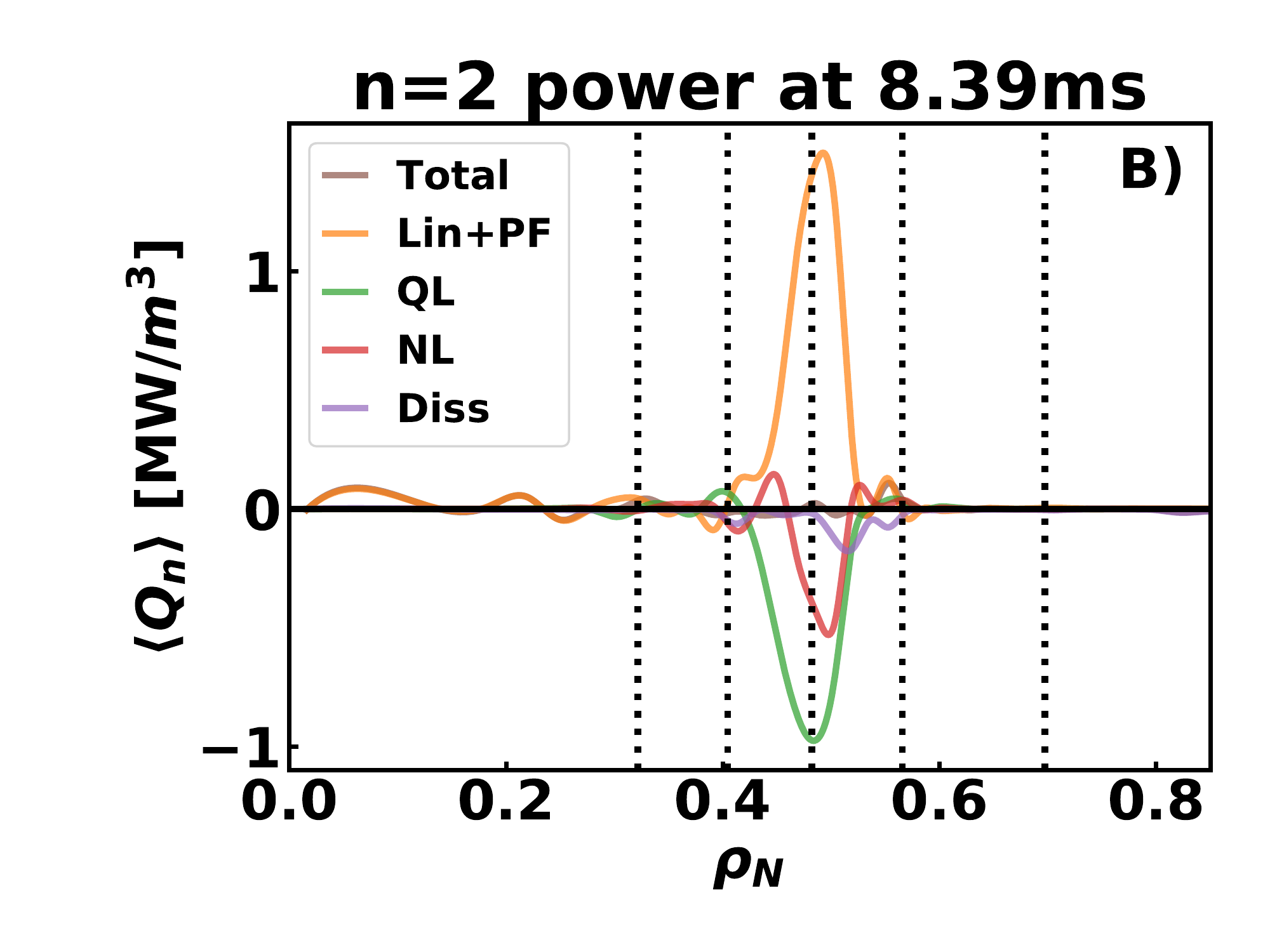}
\newline
\includegraphics[width=0.5\textwidth]{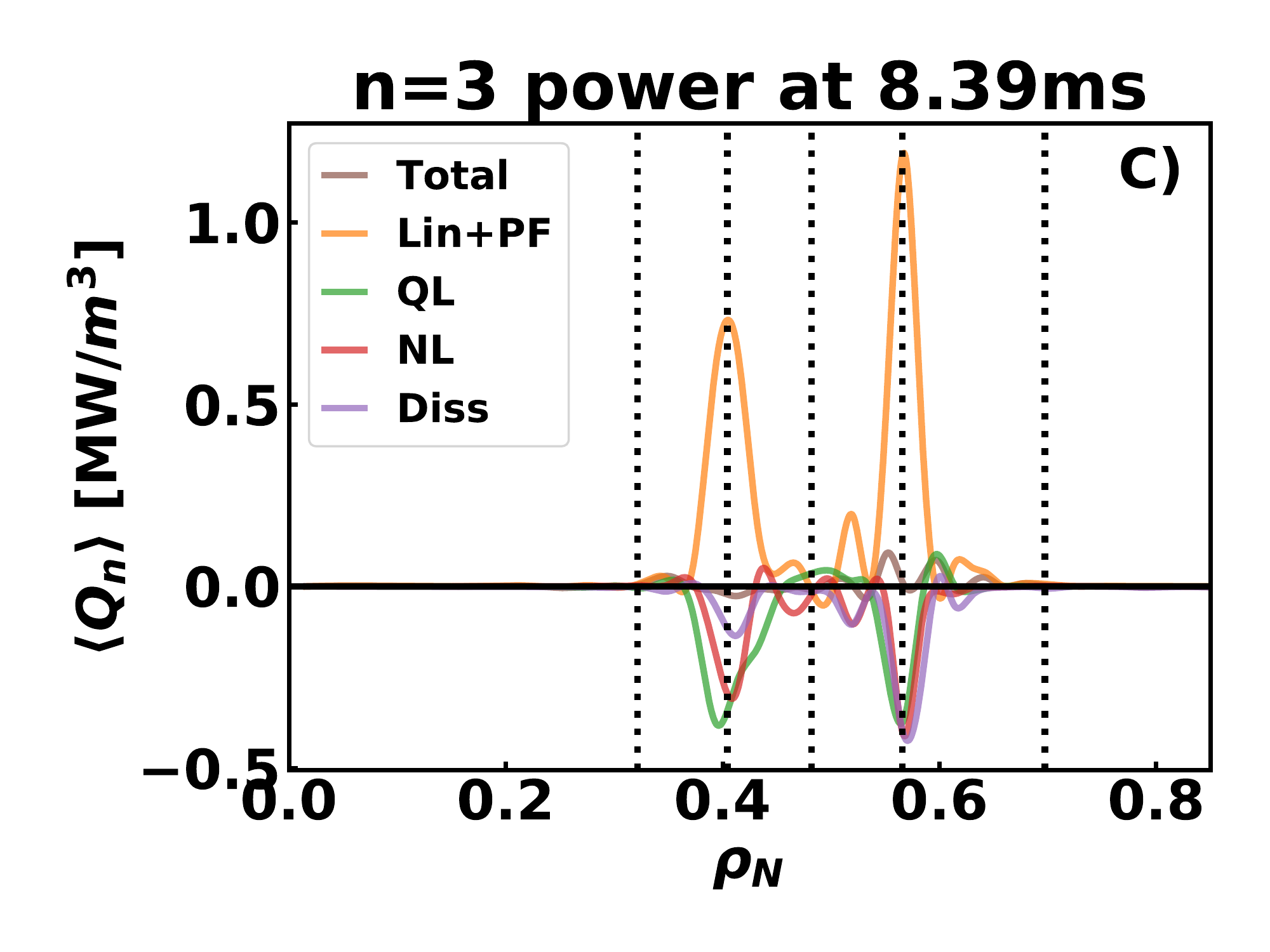}
\caption{FSA power at \SI{8.39}{ms}. A) The nonlinear $n=1$ power spans the region $\rho_N \approx 0.3$ to $\rho_N \approx 0.65$ and results from multiple nonlinear interactions (only the NL and Total powers are shown for clarity). B) The $n=2$ power is localized around the $q=1.5$ surface indicative of a $3/2$ mode. C) The $n=3$ power is localized around the $q=1.33$ and $q=1.67$ surfaces indicative of a $4/3$ and a $5/3$ tearing mode. From left to right, the dotted vertical lines indicate $q=1.2, 1.33, 1.5, 1.67,\text{ and } 2.0$ surfaces.}
\label{fig:fsa_power_8ms}
\end{figure}

The $n=1$ nonlinear power transfer at \SI{8.4}{ms} involves multiple 3-wave nonlinear interactions. Figure \ref{fig:fsa_power_8ms}A shows the nonlinear and total $n=1$ power at this time. The power is being injected into the $n=1$ across a broad radial range from $\rho_N \approx 0.3$ to $\rho_N \approx 0.65$. There are multiple peaks in the power that correlate with different low-order rational surfaces. The two largest peaks are located just inside the $q=4/3$ surface and centered on the $q=5/3$ surface. These are due to the nonlinear interactions with the $n=3$ mode. As seen in Fig. \ref{fig:fsa_power_8ms}C, the $n=3$ power is localized to two regions: one around each of these surfaces, indicative of both a $4/3$ and $5/3$ island at this time. In both regions the nonlinear $n=3$ power is negative, indicating power transfer out of the $n=3$. There is also a region of positive nonlinear power into the $n=1$ around the $q=3/2$ surface, and as seen in Fig. \ref{fig:fsa_power_8ms}B the $n=2$ nonlinear power is negative in this region.

This analysis highlights the role of 3-wave nonlinear coupling in driving the $n=1$ growth. First, a small $2/1$ island forms as a result dynamics following  the applied MP. Then at around \SI{5}{ms}, when the $6/5$ island grows large it interacts with a $7/6$ mode to transfer energy into the $n=1$. This nonlinear energy transfer helps sustain the slow growth of a $2/1$ NTM. As additional core modes grow up, they too interact nonlinearly transferring energy into the $n=1$. This process continues until the $2/1$ island surpasses a critical width; then the growth of the $2/1$ transitions to a fast growth phase that is driven by the neoclassical bootstrap current drive. A critical point is that here, the nonlinear seeding of the NTM isn't due to a single nonlinear 3-wave coupling interaction, but instead it results from a sequence of 3-wave coupling interactions that evolve in time.

%% file: discussion.tex
\section{Discussion and Conclusions} \label{sec:discussion}
This paper discusses simulations of a $2/1$ NTM seeded by a transiently applied magnetic perturbation in a tokamak plasma that approximates an ITER baseline scenario (IBS) discharge in DIII-D. The perturbation destabilizes a sequence of core modes. As these core modes grow in amplitude they nonlinearly interact depositing power into a slowly growing $2/1$. The slow growth of the $2/1$ continues until it reaches a critical amplitude, upon which it transitions to robust modified Rutherford equation growth. 

The paper also generalized the power transfer analysis of Ho and Craddock \cite{Ho1991} to shaped tokamak equilibria. This type of analysis has been useful for studying nonlinear MHD dynamics in RFPs, (e.g., Refs. \cite{Ho1991} and \cite{Reynolds08} ). Here, the power transfer analysis shows slow growth phase of $n=1$ is driven by multiple nonlinear interactions. The success here, suggests that this analysis is helpful in probing the nonlinear dynamics of low mode number magnetic perturbation in tokamaks.

These NTM dynamics are sensitive to the neoclassical poloidal flow damping rates. While not presented, we have preformed simulations varying the applied perturbation amplitude and width, and varying the neoclassical damping rates. This paper presented the results of the most interacting case, which uses an experimentally relevant damping rate. This damping rate has an intermediate value, and the nonlinear interaction between multiple core modes is a critical ingredient in the seeding of the NTM. If the electron damping rate is increased the pulse is sufficient to trigger robust growth of the $2/1$ mode. In this case an inverse sequence of modes is observed. As the $2/1$ mode grows to large amplitude, the $3/2$ mode is first excited, and then the $4/3$, $5/4$ and higher-order core modes modes are excited in sequence. In contrast if the damping rate is reduced, the plasma returns to a quiescent state following the pulse. The dynamics are less sensitive to the applied pulse under these conditions. A similar sequence of modes is observed when the pulse amplitude is decreased by 50\% or the pulse width is reduced. The pulse width and amplitude was not increased significantly to conserve computing resources.

While the goal of this study isn't to reproduce the experiment, it is interesting to compare this simulation with the experiment. In both simulation and experiment a similar sequence of core modes is observed. The magnetic signals show a $4/3$ NTM and $3/2$ NTM prior to the $2/1$. However, they evolve over different time scales. In the experiment the core modes evolve over \SI{100}{ms} time scales, and they exist prior to the ELM that triggers the $2/1$ robust growth. In the simulation, the core modes evolve on a faster few ms time scale, and they arise as a result of the applied MP. The differences in time scales may partially be due the increased resistivity in the simulation, as well as slight modifications to the core equilibrium profiles in the simulation. In both cases, the $2/1$ NTM shows two distinct phases of growth: a slow phase that is followed by a fast MRE robust growth phase. In the experiment the slow growth phase arises following an ELM, and then a subsequent ELM triggers the robust growth. In the simulations, nonlinear interactions between core modes trigger the robust growth. 

Future work will vary the initial conditions and applied MP to attempt to more accurately model the experimental observations. For example, an a applied MP with moderate $n$ ($n=6,7$) localized to the outboard mid-plane will be used to more closely model the perturbation due to an ELM. Additionally, the initial conditions can be modified to include a saturated $4/3$ and $3/2$. The power analysis can then applied to study how the moderate-$n$ interacts with preexisting islands to seed a $2/1$.

A novel application of the simulation technique developed here is to use the applied  external MP to gauge an equilibrium's resilience to NTMs. One of the challenges in experimental shot design is evaluating an equilibrium's stability to various MHD modes. NTMs are metastable states, and there is no known metric that  determines if a plasma is resilient to NTM growth. An often advocated approach calculates the classical tearing stability index $\Delta'$ \cite{GGJ}; however, the MRE analysis in Reference \cite{Callen21} shows that this term is negligible in the DIII-D IBS plasma considered here. The fact that application of a MP of sufficient amplitude in \textsc{nimrod} is required to seed both the slow and robust growth of a $2/1$ agrees with this assessment. 

Instead, it is proposed that robustness to growing NTMs could be assessed in multiple simulations by independently applying a variety of perturbations of differing amplitude, width, and mode spectrum to determine a threshold needed to excite a growing NTM. Further work is needed to refine this idea, develop an MP test suite, and then benchmark the approach with experimental data.

%% file: acknowledge.tex
\begin{acknowledgments}
E.C. Howell thanks the NTM seeding group for useful discussions on topics related to this work. This material is based on work supported by the U.S. Department of Energy Office of Science and the SciDAC Center for Extended MHD Modeling under contract numbers DE-SC0018313, DE-FC02-04ER54698, and DE-FG02-86ER53218. This research used resources of the National Energy Research Scientific Computing center, a DOE Office of Science User Facility supported by the Office of Science of the U.S. Department of Energy under contract number DE-AC02-05CH11231.

DISCLAIMER: This report was prepared as an account of work sponsored by an
agency of the United States Government. Neither the United States Government
nor any agency thereof, nor any of their employees, makes any warranty,
express or implied, or assumes any legal liability or responsibility for the
accuracy, completeness, or usefulness of any information, apparatus, product,
or process disclosed, or represents that its use would not infringe privately
owned rights. Reference herein to any specific commercial product, process,
or service by trade name, trademark, manufacturer, or otherwise, does not
necessarily constitute or imply its endorsement, recommendation, or favoring
by the United States Government or any agency thereof. The views and opinions
of authors expressed herein do not necessarily state or reflect those of the
United States Government or any agency thereof.

\end{acknowledgments}

%% file: data.tex
\section*{Data Availability}
The data that supports the findings of this study is available from the corresponding author upon reasonable request.

%% file: appendex.tex
\section{Appendix: Derivation of Power Transfer Analysis} \label{apx:PowerTransfer}

The power transfer diagnostic analysis performed in Section \ref{sec:analysis}
is discussed in detail here. The analysis generalizes the power transfer analysis of Ho and Craddock \cite{Ho1991} to toroidal geometry.

The total energy in the plasma is the sum of the magnetic, kinetic, and internal energy:
\begin{equation}
E_{total} = E_{kin} + E_{mag} + E_{int}.
\end{equation}
\noindent These three components of the energy are expressed as the volume integrals of their respective energy densities:

\begin{align}
& E_{kin} = \int dV e_{kin} = \int dV \frac{\rho v^2}{2}, \\
& E_{mag} = \int dV e_{mag} = \int dV \frac{B^2}{2\mu_0}, \\
& E_{int} = \int dV e_{int} = \int dV \frac{p}{\Gamma -1}.
\end{align}
\noindent Using flux aligned symmetric coordinates $\left(\rho, \Theta, \phi\right)$, with geometric toroidal angle $\phi$, the volume integral is
\begin{equation}
\int dV = \int d\rho \oint d\Theta \oint d \phi \mathcal{J} = \int dA \oint d \phi,
\end{equation}

\noindent where the Jacobian, $\mathcal{J}\left(\rho,\Theta \right)$, is toroidally axisymmetric.

The scalar fields $\rho$ and $p$ and the components of the vector fields $\vec v$ and $\vec B$ are projected onto a finite Fourier series. For
example, the magnetic field is expressed as:

\begin{align}
& \vec B = B_R \hat e_R + B_Z \hat e_Z + B_\phi \hat e_\phi, \\
& B_{R} = \sum_{n=-nmax}^{nmax} B_{R,n}\left(R,Z\right) e^{-in\phi}, \\
& B_{Z} = \sum_{n=-nmax}^{nmax} B_{Z,n}\left(R,Z\right) e^{-in\phi}, \\
& B_{\phi} = \sum_{n=-nmax}^{nmax} B_{\phi,n}\left(R,Z\right) e^{-in\phi},
\end{align}

\noindent where $(R,Z,\phi)$ are cylindrical coordinates. Using the notation $\vec B_n = \left(B_{R,n},B_{Z,n},B_{\phi,n}\right)$, and a similar notation for $\vec v_n$, the kinetic, magnetic, and internal energy are:

\begin{align}
&E_{kin} \approx \int dA \oint d\phi \frac{\rho_0}{2}\sum_{n,n'}\vec v_n\cdot \vec v_{n'} e^{i\left(n+n'\right)\phi}, \\ \label{eqn:KinEnergy}
&E_{mag} = \int dA \oint d\phi \sum_{n,n'}\frac{\vec B_n\cdot \vec B_{n'}}{2\mu_0} e^{i\left(n+n'\right)\phi},\\
&E_{int} = \int dA  \oint d\phi \sum_{n} \frac{p_n}{\Gamma -1}e^{i n\phi}.
\end{align}

\noindent The perturbed density, $\widetilde \rho$, is assumed to be small compared to the equilibrium density, and terms of this order are neglected in the kinetic energy. The integral over the toroidal angle, $\phi$, is exactly zero when the argument in the exponent in a nonzero integer multiple of $i\phi$. This motivates the definition of toroidal Fourier mode energy densities:

\begin{align}
& e_{kin,n} \equiv  \left\{
  \begin{aligned}
    \frac{\rho_0}{2}\vec v_n \cdot \vec v_n^* \quad \text{for } n = 0, \\
    \frac{2\rho_0}{2}\vec v_n \cdot \vec v_n^*\quad \text{for } n \neq  0,
  \end{aligned}
  \right.\\
& e_{mag,n}  \equiv \left\{
  \begin{aligned}
    \frac{1}{2\mu_0}\vec B_n \cdot \vec B_n^* \quad \text{for } n = 0, \\
    \frac{2}{2\mu_0}\vec B_n \cdot \vec B_n^*\quad \text{for } n \neq  0,
  \end{aligned}
  \right.\\
& e_{int,n}   \equiv \left\{
  \begin{aligned}
    &\frac{p_0}{\Gamma-1} &&\quad \text{for } n = 0 \\
    &0 &&\quad \text{for } n \neq 0.
  \end{aligned}
  \right.
\end{align}

\noindent The total energy is the sum over the toroidal Fourier mode energy densities:
 \begin{align}
 & E_{kin} \approx 2\pi \int dA \sum_{n=0}^{nmax} e_{kin,n}\\
 & E_{mag} = 2\pi \int dA \sum_{n=0}^{nmax} e_{mag,n} \\
 & E_{int} = 2\pi \int dA  \sum_{n=0} e_{int,n} .
 \end{align}

There are a couple of noteworthy points. First, the fact that $f_n^* = f_{-n}$ is used to change the limits of the summation from $[-\text{nmax},\text{nmax}]$ to $[0,\text{nmax}]$. This results in the factor of two in the definitions for the $n \neq 0$ toroidal mode energy densities. Second, while the total energy can be expressed as a sum of the toroidal mode energies, the same is not true for the total energy density, for example

\begin{equation}
e_{mag} \neq \sum_{n=0}^{nmax} e_{mag,n}.
\end{equation}

\noindent There are extra contributions to local energy densities that cancel in the integral over the toroidal angle. Third, the linear eigenmodes of an axisymmetric toroidal equilibrium have a discrete toroidal mode number $n$. The toroidal energy density is total energy summed over all the eigenmodes with the same $n$. However, the linear eigenmodes are composed of an infinite spectrum of poloidal Fourier modes. The Fourier energy doesn't nicely generalize to a poloidal Fourier mode energy in toroidal geometry. Finally, the internal toroidal energy density associated with the $n \neq 0$ modes is exactly zero, which greatly simplifies the following power analysis. This fact results from the choice to Fourier expand the pressure instead of the density and temperature. 

Next, the power transfer associated with the different toroidal modes is determined. The goal is to quantify the power transfer into and out of the $n \neq 0$ modes, and only the power transfer associated with the magnetic and kinetic energies needs to be considered. The power associated with $n$-th toroidal mode's magnetic energy is calculated using the induction equation:

\begin{align}
&\frac{d}{dt}e_{mag,n} = \frac{2}{2\mu_0}\frac{\partial \vec B_n}{\partial t}\cdot \vec B_n^* + c.c. \quad \text{for } n\neq 0, \\
&\frac{1}{\mu_0}\frac{\partial \vec B_n}{\partial t}\cdot \vec B_n^* = -\nabla \cdot \frac{\vec E_n \times \vec B_n^*}{\mu_0} -\vec E_n \cdot \vec J_n^*.
\end{align}

\noindent Here $c.c.$ are the complex conjugate contributions, and the electric field is calculated from Ohm's law, Eq. (\ref{eqn:ohms}). This power has two terms: the Poynting flux and the Joule power. The Poynting flux radially redistributes the magnetic energy within a toroidal Fourier mode, but it is conservative and does not transfer energy between different toroidal modes. The Joule power term is a source term, and it captures the transfer of magnetic to kinetic energy within a toroidal mode as well as the transfer of magnetic energy to different toroidal modes. 

The power associated with the $n$-th toroidal mode's kinetic energy is calculated from the momentum equation:

\begin{align}
&\frac{d}{dt}e_{kin,n} \approx \frac{2\rho_0}{2}\frac{\partial \vec v_n}{\partial t}\cdot \vec v_n^* + c.c. \quad \text{for } n\neq 0, \\
\begin{split}
&\rho_0\frac{\partial \vec v_n}{\partial t}\cdot \vec v_n^* \approx -\rho_0 \left(\vec v \cdot \nabla \vec v\right)_n \cdot \vec v_n^* \\
&+ \left(\vec J \times \vec B\right)_n \cdot \vec v_n^* - \nabla p_n\cdot \vec v_n^* + \nabla \cdot \vec {\vec {\Pi}}_n\cdot \vec v_n^*.
\end{split}
\end{align}

\noindent Here, terms of $\mathcal{O}\left( \frac{\widetilde \rho}{\rho_0}\right)$ have been neglected consistent with the definition of the kinetic energy in Eq. (\ref{eqn:KinEnergy}). The kinetic power includes an advective power, a Lorentz power, a compressive power, and a viscous dissipation power. 

The total power associated with a $n \neq 0$ mode is the sum of the powers associated with the kinetic and magnetic energies. In the paper the powers are grouped as follows:

\begin{align}
&\text{Advective}: -\rho_0 \left(\vec v \cdot \nabla \vec v\right)_n \cdot \vec v_n^*  + c.c., \label{eqn:power_adv}\\
\begin{split}
&\text{Linear}:  \left(\vec J_{eq} \times \vec B_n + \vec J_n \times \vec B_{eq}\right) \cdot \vec v_n^* \\
&\quad+\left(\vec v_{eq} \times \vec B_n + \vec v_n \times \vec B_{eq}\right) \cdot \vec J_n^* \\
&\quad- \nabla p_n\cdot \vec v_n^* + c.c.,
\end{split}\\
\begin{split}
&\text{Quasi-linear}: \left(\widetilde J_{0} \times \vec B_n + \vec J_n \times \widetilde B_{0}\right) \cdot \vec v_n^* +\\
&\quad \left(\widetilde v_{0} \times \vec B_n + \vec v_n \times \widetilde B_{0}\right) \cdot \vec J_n^* + c.c., \\
\end{split}\\
\begin{split}
&\text{Nonlinear}: \left(\widetilde J \times \widetilde B\right)_n \cdot \vec v_n^* +\\
&\quad \left(\widetilde v \times \widetilde B\right)_n \cdot \vec J_n^* + c.c., \\
\end{split}\\
\begin{split}
&\text{Dissipative}:-\eta J_n^2 +\frac{1}{n_0e}\nabla \cdot \vec {\vec {\Pi}}_{e,n}\cdot \vec J_n^*+\\
&\quad\nabla \cdot \vec {\vec {\Pi}}_{i,n}\cdot \vec v_n^* + c.c.,\\
\end{split}\\
&\text{Poynting Flux}: -\nabla \cdot \frac{\vec E_n \times \vec B_n^*}{\mu_0} + c.c..\label{eqn:power_pf}
\end{align}

\noindent Here the $\widetilde f$ represents the perturbed quantities, and the triple products in the nonlinear terms neglect both equilibrium and perturbed $n=0$ contributions to avoid double counting.

The linear terms are the terms present in an ideal MHD analysis of a static equilibrium (the Poynting flux also has a term that contributes to the ideal MHD analysis). The linear terms represent a transfer of energy between the $n$-th toroidal mode and the initial equilibrium fields. Since the energy in the perturbations is small compared to the energy in the equilibrium, the linear terms are expected to have a large contribution to the power balance even late into the nonlinear evolution. The quasi-linear terms represent the power transfer between $n=0$ modifications to the equilibrium and the $n$-th toroidal mode. The terms quantify how quasi-linear changes to the equilibrium, such has flattening of the pressure profile, drive the $n$-th toroidal mode. The nonlinear terms quantify the power transfer between two non-axisymmetric toroidal modes. 

The Poynting flux represents a radial redistribution of magnetic energy within a toroidal mode. In principle it could be decomposed into linear, quasilinear, nonlinear, and dissipative terms. However, this flux doesn't involve an exchange of energy between Fourier modes and the decomposition doesn't add additional insight here, so we find it most insightful to express the Poynting flux as one term. 

The advective power could similarly be decomposed into three terms, but the advective power is small. The power is included in the calculation of the total power for completeness, but doesn't meaningfully impact the power here.

The dissipative terms group the powers associated with the resistivity, neoclassical electron stress, and total ion viscosity. The terms associated with the resistivity and neoclassical electron stress are negative. These terms enable the growth of both classical and neoclassical tearing modes. This growth is enabled not by the power transfer associated with these terms, but instead by modifications to the mode structure that arise when including these effects. 

The paper considers both the total volume integrated toroidal mode power
\begin{equation}
P_n = \frac{d}{dt}E_n = 2\pi \int \rho \oint d\Theta \mathcal{J}\frac{d}{dt}e_n
\end{equation}
and the flux-surface-averaged toroidal mode power
\begin{equation}
\langle Q_n \rangle = \left< \frac{d}{dt}e_n\right> =  \frac{1}{V'} \oint d\Theta \mathcal{J}\frac{d}{dt}e_n.
\end{equation}
The total volume integrated toroidal mode power determines the growth of a mode. A positive power indicates that the energy in the $n$-th toroidal mode is increasing and that the mode is growing. A negative power indicates the energy in the toroidal mode is decreasing and that the mode is decaying. The flux-surface-averaged (FSA) toroidal mode power provides insight into the radial power deposition. 